\RequirePackage{xcolor}
\documentclass[conference]{IEEEtran}
\usepackage{cite}
\usepackage{amsmath,amssymb,amsfonts,amsthm}
\usepackage{graphicx}
\usepackage[font=footnotesize]{caption}
\usepackage{subcaption}
\usepackage{enumerate}
\usepackage{hyperref}
\usepackage{makecell}
\usepackage{multirow}
\usepackage{booktabs}
\usepackage{bm}  
\usepackage{lipsum}
\usepackage[ruled, linesnumbered]{algorithm2e}
\usepackage{algpseudocode}
\usepackage{mathtools}
\usepackage{textcomp}
\usepackage{verbatim}
\usepackage{breqn}
\usepackage{graphicx}
\usepackage[textsize=tiny]{todonotes}
\usepackage{nomencl}

\makeatletter 
\g@addto@macro{\@algocf@init}{\SetKwInOut{Parameter}{Parameters}}

\makeatother

\newtheorem{definition}{Definition}
\newtheorem{assumption}{Assumption}
\newtheorem{lemma}{Lemma}
\newtheorem{theorem}{Theorem}
\newtheorem{remark}{Remark}

\newtheorem{problem}{Problem}
\makenomenclature

\def\BigRoman{\uppercase\expandafter{\romannumeral\number\count 255 }}
\def\Romannumeral{\afterassignment\LowRoman\count255=}
\newcommand{\RN}[1]{%
  \textup{\uppercase\expandafter{\romannumeral#1}}%
}

\usetikzlibrary{shapes,arrows}
\def\BibTeX{{\rm B\kern-.05em{\sc i\kern-.025em b}\kern-.08em
    T\kern-.1667em\lower.7ex\hbox{E}\kern-.125emX}}
\begin{document}
\title{An Invariant Set Construction Method, Applied to Safe Coordination of Thermostatic Loads}
\author{Sunho Jang, Necmiye Ozay, and Johanna L. Mathieu
\thanks{The authors are with the Department of Electrical Engineering and Computer Science, University of Michigan, Ann Arbor, MI, 48109 USA (email: \{sunhoj, necmiye, jlmath\}@umich.edu). This work was supported by U.S. National Science Foundation Award CNS-1837680.}
}
\maketitle
\begin{abstract}
We consider the problem of coordinating a collection of switched subsystems under both local and global constraints for safe operation of the system. Although an invariant set can be leveraged to construct a safety-guaranteed controller for this kind of problem, computing an invariant set is not scalable to high-dimensional systems. In this paper, we introduce a strategy to obtain an implicit representation of a controlled invariant set for a collection of switched subsystems, and construct a safety-guaranteed controller to coordinate the subsystems using the representation. Specifically, we incorporate the invariant set into a model predictive controller to guarantee safety and recursive feasibility. Since the amount of computations is independent of the number of subsystems, this approach scales to large collections of switched subsystems. We use our approach to safely control a collection of thermostatically controlled loads to provide grid balancing services. The problem includes constraints on each load's temperature and duration it must remain in a mode after a switch, and also on aggregate power consumption to ensure network safety. Numerical simulations show that the proposed approach outperforms benchmark strategies in terms of safety and recursive feasibility.
\end{abstract}


\section{Introduction}
\label{sec:introduction}
The safety of cyber-physical systems that consist of multiple subsystems is often described by multiple levels of constraints; there are both local constraints on each subsystem and global constraints on their collective behavior. In order to guarantee safety, we can use an algorithm that keeps the state inside a controlled invariant set~\cite{bertsekas1972infinite,aubin2011viability,blanchini1999set}. Some previous work~\cite{koller2018learning,wabersich2018linear,wabersich2021probabilistic} has incorporated invariant sets into Model Predictive Control (MPC), resulting in controllers with both safety and recursive feasibility guarantees.
 
However, these methods of invariant set computation are not scalable; obtaining an explicit form of an invariant set requires a projection from a higher-dimensional state space to a lower-dimensional one, which is computationally burdensome. Some recent studies~\cite{anevlavix2019computing,fiacchini2018computing}, propose scalable algorithms to compute invariant sets for linear systems. However, these algorithms are neither scalable enough for high-dimensional systems nor applicable to switched systems, which have discrete inputs or modes. The method introduced in \cite{wintenberg2020implicit} computes an invariant set for a switched linear system; however, it is also not sufficiently scalable to deal with high-dimensional systems.

In this work, we develop a scalable method to construct a controlled invariant set for a large collection of switched subsystems, and synthesize a control algorithm with formal safety guarantees by incorporating the invariant set. We develop an approach to obtain an \textit{implicit representation} of an invariant set of a high-dimensional system and incorporate that representation into an MPC algorithm to provide guarantees on safety and recursive feasibility. Crucially, this strategy allows us to avoid the heavy computational burden imposed by the projection step so that our approach is applicable to a high-dimensional system.

Our problem is motivated by the application-domain problem of coordinating a collection of hundreds or thousands of Thermostatically Controlled Loads (TCLs) to balance supply and demand on the grid. For safety, the controller should be designed to enforce not only local constraints on each TCL but also network-level safety constraints. Many previous papers~\cite{mathieu2012state,bashash2012modeling, zhang2013aggregated,tindemans2015decentralized,coffman2019aggregate,ziras2018new} have proposed approaches that satisfy the local TCL constraints; but the safety of the distribution network (e.g., voltages maintained within their limits) is not generally considered. Some work~\cite{dall2017optimal,bernstein2019real,vrettos2013combined} deals with network safety by incorporating network-level constraints using optimal power flow approaches; but recursive feasibility of the optimization problem is not guaranteed. In contrast, our approach explicitly ensures recursive feasibility, and therefore safety. 

The main contributions of our paper are threefold. First, we develop a method for finding an implicit representation of an invariant set of a system composed of a large number of heterogeneous subsystems with global constraints on their collective behavior and local constraints including {\em lockout constraints}, which require a subsystem to keep its current mode for a certain time duration after a switch. Second, we propose a control algorithm with safety guarantees by incorporating the implicit representation of the invariant set. Third, we apply the proposed approach to the coordination of TCLs for frequency regulation while limiting aggregate power consumption to ensure network safety. We use a distribution network model to demonstrate how our controller avoids voltage violations. This paper significantly extends our preliminary work \cite{jang2021large}, which only considered homogeneous subsystems without lockout constraints and did not explicitly model the network. Here we benchmark our results against several other control approaches including the approach we proposed in~\cite{jang2021large} to demonstrate the importance of modeling lockout. 

The organization of the paper is as follows. In Section~\ref{sec:Problem_Setting}, we describe the application-domain problem and, in Section~\ref{sec:abstraction}, we detail the abstraction and aggregate system construction. In Section~\ref{sec:invsetmethod}, we propose our novel invariant set construction method and, in Section~\ref{sec:INVMPC}, our safety-guaranteed control algorithm. Numerical simulation results are given in Section~\ref{sec:simulation}. Proofs are presented in the Appendix.

\emph{Notation:}
  We write row vectors of ones as $\bm{1}$. The $l$th element of matrix $C$ is denoted $[C]_{l}$. We denote the set of non-negative integers as $\mathbb{N}_0 := \mathbb{N} \cup \{ 0 \} $. Also, $ [ X ] $ refers to the set of the integers $\{ 1, \ldots, X \}$, and $[X]_0$ denotes $[X] \cup \{0\}$. Floor and ceiling functions are represented by $\lfloor \cdot \rfloor$ and $\lceil \cdot \rceil$, respectively. The indicator function on set $A$ is denoted $\bm{1}_{A}$. 
  We denote the Minkowski sum by $A \oplus B = \{ a+b \enspace | \enspace  a \in A, b \in B \} $, and the subtraction $A \ominus B$ is defined as the largest solution to $X \oplus B = A$. Also, $\| \cdot \|$ refers to the infinity norm. The ball with radius $r$ centered at $\theta$ is denoted $\mathcal{B}(\theta,r) := \{ x \enspace | \enspace \| x - \theta \| \leq r \}$.  The identity function in space $\mathbb{R}^d$ is denoted $\text{Id}_{\mathbb{R}^d}$.
\vspace{-0.2cm}

\section{Application-domain Problem Setting} \label{sec:Problem_Setting}

We consider a power reference tracking control problem for a collection of TCLs, such as air conditioners, which switch on/off to maintain a temperature within a dead-band, i.e., a small range around a setpoint. Each TCL is a switched subsystem, which collectively make up a system. The control problem is to switch on/off individual TCLs to cause the power consumption of the collection to track a signal, e.g., a scaled and shifted frequency regulation signal, while ensuring local constraints are satisfied. First, the temperature of each TCL should remain within its dead-band. Second, the on/off mode of each TCL should be maintained for a certain duration after a switch to ensure the compressor is not damaged; this constraint is called a lockout constraint~\cite{ziras2018new}. There are several approaches that have been developed to handle lockout, e.g., \cite{coffman2019aggregate}, though to the best of our knowledge none ensures recursive feasibility and safety.

Manipulating the aggregate power consumption of TCLs could cause constraint violations in the distribution network, e.g., over/under-voltages and transformer overloading~\cite{ross_effects_2019}. One way to avoid these violations is to impose minimum and maximum bounds on the aggregate power consumption of the collection of TCLs, i.e., network-level safety constraints. A variety of recent papers have developed approaches to compute bounds on network-safe changes in power consumption/production of distributed energy resources~\cite{ross2020method, nazir2019convex,molzahn2019grid,cui2021network}. Assuming we can compute conservative bounds with one of these methods, controlling TCLs such that their aggregate power consumption remains within these bounds should guarantee network safety. Therefore, the collection of TCLs should commit to providing a regulation capacity (i.e., the range over which it can manipulate its aggregate power consumption) within the network-safe bounds. However, if the TCLs prioritize maintaining temperatures within dead-bands, attempting to track an aggressive signal within those bounds (e.g., a signal that stays at either bound for a long duration) can eventually render the signal untrackable and can cause the TCL aggregate power consumption to violate the bounds. 

\begin{figure}[t]
    \centering
    \includegraphics[width = 0.45\textwidth]{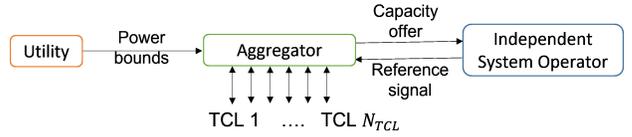}
    \caption{Interactions between the aggregator, utility, and ISO.}
    \label{fig:entitiesrelationship}
    \vspace{-0.7cm}
\end{figure}

To clarify this issue, we must first describe our assumed control architecture. There are many possible architectures for network-safe TCL participation in electricity markets~\cite{ross2019coordination},
but here we consider one in which a third-party aggregator coordinates TCLs subject to network-safe power bounds provided by the utility. Fig.~\ref{fig:entitiesrelationship} shows the interaction of the aggregator, utility, and Independent System Operator (ISO). Based on the forecast of the network states, the utility computes the range of TCL power consumption that can be safely accommodated by the network, and sends it to the aggregator. The bounds are used by the aggregator to make an offer for regulation capacity which is expected to be safe. The ISO commits the aggregator at or below its offered capacity. In real time, the ISO generates a normalized regulation signal (ranging from -1 to 1, with 0 corresponding to a resource's scheduled or nominal operating point), and the aggregator i) scales it by its committed capacity, ii) shifts it by the TCL collection's nominal load, and iii) controls the modes of TCLs to track it. Some U.S.~ISOs have designed mechanisms to constrain regulation signals to reduce the chance of aggressive signals in an effort to support the use of energy-constrained resources like energy storage and TCLs \cite{pjm2017implementation,caiso2014non}. However, many ISOs do not have these mechanisms, and so a TCL collection attempting to track an aggressive signal with a simple tracking controller may violate the network-safe power bounds. 

\begin{figure}
    \centering
    \includegraphics[width=0.35\textwidth]{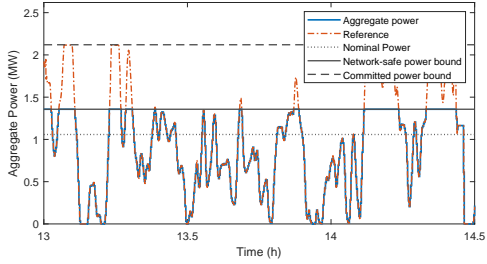}
    \caption{Reference tracking with aggregate power constraints. The reference signal (from \cite{PJMrefsignal}) is within the committed power bounds, but outside the actual network-safe power bounds. The desired power is the blue signal.}
    \label{fig:trackingex}
    \vspace{-0.6cm}
\end{figure}

This is even more likely to happen if real-time network states differ from their forecasts and the actual network-safe bounds are tighter than those generated with the forecasts. Fig.~\ref{fig:trackingex} provides an example of this situation, where the dashed black lines are the power bounds obtained in advance that are used to determine the committed capacity, and the solid black lines are the actual network-safe power bounds computed in real-time. The red reference signal (a scaled and shifted frequency regulation signal from the PJM Interconnection~\cite{PJMrefsignal}) is scaled to the committed power bounds and is impossible to track while ensuring network safety. We assume that network safety takes precedence over tracking performance, specifically, the aggregator should track the regulation signal as well as possible within the tighter bounds, i.e., the blue signal in Fig.~\ref{fig:trackingex}. PJM uses a mechanism to constrain the integral of the normalized signal (i.e., the energy content of the signal), meaning that tracking the red signal is unlikely to cause power bound violations. However, the bound tightening in Fig.~\ref{fig:trackingex} makes the blue signal much harder to track; attempting to track it with a simple tracking controller would cause average aggregate power consumption to decrease from nominal, eventually rendering the signal untrackable as TCLs struggle to cool/heat sufficiently. Eventually, many TCLs may need to be on simultaneously violating the network-safe power bound. We will benchmark our approach against a simple tracking controller in Section~\ref{sec:INVMPC}.

We next detail the problem formulation. We consider a collection of TCLs that are clustered into multiple groups based on their similarity in terms of thermal/electrical parameters and ambient temperatures, and we treat TCLs in the same group as homogeneous. To do this, the amount of deviations allowed between TCLs in the same group is small enough so that we can construct a bisimilar abstraction for each group; details are given in Section~\ref{subsec:abstractions}. Our method also allows for sufficiently small process noise and variation in ambient temperature, but our mathematical development assumes no process noise and constant ambient temperature to simplify the exposition of the main results. Considering multiple groups allows us to deal with TCL heterogeneity, which we did not consider in our preliminary work \cite{jang2021large}.

Let the number of groups of homogeneous TCLs be $g_{\text{TCL}}$, the total number of TCLs across all groups be $N_{\text{TCL}}$, and the number of TCLs in group $i$ be $N_{\text{TCL}}^{(i)}$; hence, $ \sum_{i=1}^{g_{\text{TCL}}} N_{\text{TCL}}^{(i)} = N_{\text{TCL}}$. The temperature of the $j$th TCL in group $i$ at time $t$ is denoted by $T_{j}^{(i)} (t)$, and its domain is $[T_{\text{min}}^{(i)}, T_{\text{max}}^{(i)}]$.
Also, the on/off mode of each TCL is $\mu_{j}^{(i)} (t)$, which is $0$ if it is not consuming power (off), and $1$ if it is consuming power (on). Then, we assume that the temperature evolution follows the affine model developed in \cite{sonderegger1978dynamic} as follows 
\begin{equation} \label{eq:TCLdynamics}
        T_{j}^{(i)} (t+1) = a^{(i)} T_{j}^{(i)} (t) + \left ( 1 - a^{(i)} \right ) \left ( T_{a}^{(i)}- R^{(i)} p_{\text{tr}}^{(i)} \mu_{j}^{(i)} (t)  \right ). 
\end{equation}
Here, $T_{a}^{(i)}$ is the ambient temperature of TCLs in group $i$ and $a^{(i)}$ is  $\text{exp}(-\Delta t / (R^{ (i) } C^{ (i) }))$, where $\Delta t$ is the sampling time, $R^{(i)}$ is the thermal resistance and $C^{(i)}$ the thermal capacitance of every TCL in group $i$. Parameter $p_{\text{tr}}^{(i)}$ is the thermal energy transfer rate, which is positive for a cooling TCL and negative for a heating TCL; since here we assume all TCLs are air conditioners, the value is positive. The power consumption of each TCL in group $i$ when it is in on mode, denoted $p^{(i)}$, equals $p_{\text{tr}}^{(i)} / \zeta^{(i)}$, where $\zeta^{(i)}$ is the coefficient of performance. Then, the aggregate power consumption of all TCLs, denoted $P_{\text{agg}}(t)$, equals $\sum_{i=1}^{g_{\text{TCL}}} p^{(i)} \sum_{j=1}^{N_{\text{TCL}}^{(i)}} \mu_{j}^{(i)} (t)$. 
        
We next define the local constraints. The temperature dead-band of group $i$ is $[\underline{T}^{(i)}, \overline{T}^{(i)}]$ ($\overline{T}^{(i)} \leq T_{\text{max}}^{(i)}, \, \underline{T}^{(i)} \geq T_{\text{min}}^{(i)}$), which leads to the constraint 
\begin{equation} \label{eq:TCLlocalconstraint}
        T_{j}^{(i)} (t) \in [ \underline{T}^{(i)}, \overline{T}^{(i)} ] \quad \forall t \in \mathbb{N}_0, i \in [g_{\text{TCL}}], j \in [N_{\text{TCL}}^{(i)}].
\end{equation}
We assume that temperature setpoints and dead-bands are constant over time.
 
Additionally, a lockout constraint on every TCL requires it to remain in the same mode for a specific duration after a switch; imposing this constraint is a key extension from our preliminary work~\cite{jang2021large}. Suppose that the mode of every TCL in group $i$ should be kept for $ \overline{t}_{\text{off}}^{(i)}$ after it is turned off and for $ \overline{t}_{\text{on}}^{(i)}$ after it is turned on. Then, the number of time steps TCL should be locked after a switch for each mode is computed as 
\begin{equation*}
    \overline{\tau}_{\text{off}}^{(i)} = \left \lceil \frac{\overline{t}_{\text{off}}^{(i)}}{\Delta t} \right \rceil, \quad \overline{\tau}_{\text{on}}^{(i)} = \left \lceil \frac{ \overline{t}_{\text{on}}^{(i)}}{\Delta t} \right \rceil.
\end{equation*}
Now, let $\tilde{t}_{j,\text{off}}^{(i)} (t)$ / $\tilde{t}_{j,\text{on}}^{(i)} (t) $ be the last time step the $j$th TCL in group $i$ switched its mode to off/on. Then, the input sequence $\mu_j^{(i)} (t)$ should satisfy the following 
\begin{equation} \label{eq:lockoutconstraints}
    \begin{aligned}
        & \mu_j^{(i)} (t) = 0  \quad (\text{if } t < \tilde{t}_{j,\text{off}}^{(i)} (t) + \overline{\tau}_{\text{off}}^{(i)} ) \text{ and} \\
        & \mu_j^{(i)} (t) = 1  \quad (\text{if } t < \tilde{t}_{j,\text{on}}^{(i)} (t) + \overline{\tau}_{\text{on}}^{(i)} ).
    \end{aligned}
\end{equation}

Next, we define the global constraint, which ensures the safety of the distribution network. Specifically, we bound the aggregate power consumption
 \begin{equation} \label{eq:TCLaggconstraint}
    \underline{P}_{\text{agg}} \leq P_{\text{agg}} (t) =  \sum_{i=1}^{g_{\text{TCL}}} p^{(i)} \sum_{j=1}^{N_{\text{TCL}}^{(i)}} \mu_{j}^{(i)} (t) \leq \overline{P}_{\text{agg}} \quad \forall t \in \mathbb{N}_0,
\end{equation}
where $\overline{P}_{\text{agg}},\underline{P}_{\text{agg}}$ are the upper and lower power bounds.

Finally, our goal is to control the modes of the TCLs so that the aggregate power consumption $P_{\text{agg}}(t)$ of all TCLs tracks the reference signal $r(t)$ as well as possible while satisfying all constraints. Then, the problem of interest is as follows.
\begin{problem} \label{pr:TCLproblem}
    Given the system described above, synthesize a controller to choose $\mu_{j}^{(i)} (t)$ for all $i=1,\ldots,g_{\text{TCL}}$, $j=1,\ldots,N_{\text{TCL}}^{(i)}$, and $t \in \mathbb{N}_0$ that guarantees satisfaction of constraints \eqref{eq:TCLlocalconstraint}, \eqref{eq:lockoutconstraints}, \eqref{eq:TCLaggconstraint} while trying to minimize the performance measure $| P_{\text{agg}}(t) - r(t) |$.
\end{problem}

\section{Abstraction and aggregation of a collection of switched subsystems} \label{sec:abstraction}
 In this section, we consider an abstract problem where the goal is to coordinate a collection of switched subsystems, clustered into groups based on the similarity of their dynamics, subject to the constraints that are generalized versions of \eqref{eq:TCLlocalconstraint}, \eqref{eq:lockoutconstraints}, and \eqref{eq:TCLaggconstraint}.
 After we introduce this general problem setup, we explain how to construct an abstraction with discrete-states from the original continuous-state subsystems in each group. This abstraction is then used to model an ``aggregate system", representing the dynamics, which we use to synthesize a safety-guaranteed controller.
 
 Using aggregate dynamics to represent collections of TCLs is not a new idea \cite{mathieu2012state,bashash2012modeling, zhang2013aggregated} but works that aim to characterize the relation between the aggregate dynamics and the actual collection are more recent. Ref.~\cite{soudjani2014aggregation} abstracts a TCL as a Markov chain and probabilistically quantifies the difference between the dynamics of the underlying collection and its aggregate model. In contrast, our approach provides deterministic bounds using approximate bisimulation relations. Furthermore, their model is different because the mode of each TCL is assumed to be uncontrollable; instead their input is the temperature setpoint. Ref.~\cite{nilsson2019control} uses a similar approach to ours to construct abstractions and an aggregate system, but it considers continuous-time subsystems and develops open-loop control signals, whereas we consider discrete-time subsystems and propose an algorithm to compute invariant sets, which can be used to ensure safety of arbitrary feedback controllers. The abstraction and aggregation methods from our preliminary work \cite{jang2021large} are only applicable to collections of homogeneous subsystems, whereas here we consider heterogeneous subsystems. Additionally, none of the above-mentioned approaches consider lockout constraints, as we do here.
 
 \subsection{Abstract problem formulation} \label{sec:system_model}
   
   First, the \textit{transition systems} formalism is introduced to model discrete-time dynamics~\cite{tabuada2009verification}. 
    \begin{definition} \label{def:transsys}
        A transition system $T$ is a tuple $(X,U,\rightarrow,Y)$, where $X$ is a set of states, $U$ a set of actions, $\rightarrow \subset X \times U \times X$ a transition relation, and $Y : X \rightarrow \mathbb{R}^n$ an output function. 
    \end{definition}
    
    We denote $(x,u,x') \in \rightarrow$ as $x \xrightarrow[]{u} x'$  for short.
    
    \begin{definition} \label{def:CIS}
        Given a transition system $T=(X,U,\rightarrow,Y)$, a safe set $X_{\text{safe}}\subset X$, and set of admissible input constraints $U_{\text{safe}}\subset U$, a set $X_{\text{inv}}$ is a controlled invariant set with respect to $(T,X_{\text{safe}},U_{\text{safe}})$ if $X_{\text{inv}}\subset X_{\text{safe}}$ and for all $x\in X_{\text{inv}}$ there exists $u\in U_{\text{safe}}$ such that for all $x'$ with $x \xrightarrow[]{u} x'$, we have $x'\in X_{\text{inv}}$. The union of all controlled invariant sets with respect to $(T,X_{\text{safe}},U_{\text{safe}})$ is called the maximal controlled invariant set for $(T,X_{\text{safe}},U_{\text{safe}})$.
    \end{definition}
    
We consider a system that includes multiple groups of homogeneous switched subsystems. Here, the number of groups is $g$, the number of subsystems in group $i$ is $N^{(i)}$, the total number of subsystems across all groups is $N$ (i.e., $N = \sum_{i=1}^{g} N^{(i)}$), and the number of modes is $M$.
     
 For any $i \in [g ]$ and $j \in [ N^{(i)} ]$, let $S_j^{(i)}$ be the $j$th subsystem of group $i$, and $\theta_{j}^{ (i) } (t) \in \Theta^{(i)}$ be its state at time step $t$, where $\Theta^{ (i) } \subset \mathbb{R}^d$ is a compact domain. The difference equation governing the evolution of $\theta_{j}^{(i)} (t)$ is
    \begin{equation} \label{eq:dynoforig}
        S_j^{ (i) } : \enspace \theta_{j}^{(i)} (t+1) = f^{(i)} \left(\theta_{j}^{(i)} (t), \mu_{j}^{(i)} (t)\right), \quad \mu_j^{(i)} : \mathbb{N}_0 \rightarrow [M],
    \end{equation}
    where $\mu_j^{(i)} (t)$ is the mode of $j$th subsystem in group $i$ at time $t$ and is also the control input. For TCLs, $\theta_{j}^{(i)}$ corresponds to $T_{j}^{(i)}$ and $f^{(i)} $ corresponds to the affine dynamics in \eqref{eq:TCLdynamics}. For the rest of this section, we drop the index $j$ for convenience.
    
    The discrete-time dynamics $S^{(i)}$ in \eqref{eq:dynoforig} can be equivalently represented as a transition system 
     \begin{equation} \label{eq:trnsrepoforig}
        S^{(i)} = \left(\Theta^{(i)}, [M], \xrightarrow[(i)]{}, \text{Id}_{\mathbb{R}^d}\right),
    \end{equation}
    where $\theta \xrightarrow[(i)]{\mu} \theta'$ if and only if $\theta' = f^{(i)} (\theta,\mu)$.
    
    To formally describe each group of subsystems, we introduce the following definition of a product transition system.
    \begin{definition} \label{def:prod_transsys}
        Given $N$ identical copies of a transition system $T=(X,U,\rightarrow,Y)$, the product transition system is given by $T^{\times N}=(X^N,U^N,\xrightarrow[\times N]{},Y^{\times N})$, where $(x_1,\ldots, x_N) \xrightarrow[\times N]{(u_1,\ldots, u_N)}{}(x'_1,\ldots, x'_N)$
        if and only if $x_i\xrightarrow[]{u_i}x'_i$ for all $i\in[N]$; and $Y^{\times N}:(x_1,\ldots, x_N)\mapsto  (Y(x_1),\ldots, Y(x_N))$.
    \end{definition}
    
   According to the definition above, group $i$ can be represented by the product transition system  $ S^{(i) \times N^{(i)} }$. Furthermore, the entire system can be represented as the product of those systems, denoted 
   $ \bm{S} :=  ( S^{(1) \times N^{(1)}} ) \times \cdots \times ( S^{( g ) \times N^{(g)}})$ with the set of states $\bm{\Theta} := (\Theta^{(1)})^{N^{(1)}} \times \cdots \times (\Theta^{(g)})^{N^{(g)}}$. 

 Now, we introduce the constraints imposed on $\bm{S}$. First, the state $\theta_j^{(i)}$ of each subsystem in group $i$ should stay in the safe set $\Theta_{\text{safe}}^{(i)} $ (which corresponds to \eqref{eq:TCLlocalconstraint} in the TCL problem), i.e.,
\begin{equation} \label{eq:generalsafesetcond}
    \theta_j^{(i)} (t) \in \Theta_{\text{safe}}^{(i)} \quad \forall j \in [ N^{(i)} ], t \in \mathbb{N}_0.
\end{equation}
Second, each subsystem in group $i$ cannot switch again for $\overline{\tau}_{m}^{(i)}$ time steps after its mode is switched to $m$ (which corresponds to the lockout constraints  \eqref{eq:lockoutconstraints} in the TCL problem), and so the input sequence $\mu_j^{(i)} (t)$ of each subsystem should satisfy
\begin{equation} \label{eq:generallockout}
\mu_j^{(i)}(t) =m \enspace \text{if} \enspace t < \tilde{t}_{j,m}^{(i)} (t) + \overline{\tau}_m^{(i)} \quad \forall m \in [M],
\end{equation}
where $\tilde{t}_{j,m}^{(i)} (t)$ is the last time step the subsystem $S_j^{(i)}$ switched its mode to $m$ (i.e., $\mu(\tilde{t}_{j,m}^{(i)}) = m$, $\mu(\tilde{t}_{j,m}^{(i)}-1) \neq m$). Third and finally, we impose bounds on a linear combination of the number of subsystems in each mode $m$ in each group $i$ (which corresponds to \eqref{eq:TCLaggconstraint} in the TCL problem) as follows
\begin{equation} \label{eq:generalaggconstraints}
    \begin{aligned}
         \underline{P}_m \leq \sum_{i=1}^{g}  p_{m}^{(i)} \sum_{j=1}^{N^{(i)}} \bm{1}_{m} \left(\mu_{j}^{(i)} (t)\right) \leq \overline{P}_m \enspace \forall & m \in [M], t \in \mathbb{N}_0,
    \end{aligned}    
\end{equation} 
 where $p_{m}^{(i)}$ is a scalar corresponding to subsystems in mode $m$ in group $i$, and $\sum_{j=1}^{N^{(i)}} \bm{1}_{m} \left(\mu_{j}^{(i)} (t)\right)$ is the number of subsystems in mode $m$ in group $i$.

Additionally, we assume a time-varying cost function $c_t : \mathbb{N}_0^{gM} \rightarrow \mathbb{R}$ which depends on the number of subsystems in each mode in each group.
Then, the generalized problem is as follows.
\begin{problem} \label{pr:generalproblem}
    Given a collection of subsystems $S_j^{(i)}$ as described above, synthesize a controller to choose $\mu_{j}^{(i)} (t)$ for all $i =1,\ldots,g$, $j=1,\ldots,N^{(i)}$, and $t \in \mathbb{N}_0$ that guarantees satisfaction of constraints \eqref{eq:generalsafesetcond}, \eqref{eq:generallockout}, \eqref{eq:generalaggconstraints} while trying to minimize a cost function $c_t$.
\end{problem}

\subsection{Abstractions} \label{subsec:abstractions}

   In this section, we introduce the abstraction of each subsystem $S_j^{(i)}$. We consider the following notion of closeness to the original subsystem.
     \begin{definition} \label{def:epsbisimilar}
        Two transition systems $T_1=(X_1,U, \xrightarrow[1]{}, Y_1)$ and $T_2=(X_2,U, \xrightarrow[2]{}, Y_2)$ are $\epsilon$-approximately bisimilar if there exists a relation $R \subset X_1 \times X_2$ such that the sets $\hat{R}_{1 \rightarrow 2} = \{ x_2 : (x_1, x_2) \in R \}$ and $\hat{R}_{2 \rightarrow 1} (x_2) = \{ x_1 : (x_1,x_2) \in R \}$ are non-empty for all $x_1,x_2$, and such that for all $(x_1,x_2) \in R$, all of the following are satisfied.
        \begin{itemize}
            \item $\| Y_1(x_1) - Y_2(x_2) \|  \leq \epsilon $.
            \item If $x_1 \xrightarrow[1]{u} x_1'$, there exists $x_2 \xrightarrow[2]{u} x_2'$ s.t. $(x_1',x_2') \in R$.
            \item If $x_2 \xrightarrow[2]{u} x_2'$, there exists $x_1 \xrightarrow[1]{u} x_1'$ s.t. $(x_1',x_2') \in R$.
        \end{itemize}
    \end{definition}

To guarantee the existence of a bisimilar abstraction for the subsystems in each group, the following assumption is made.
\begin{assumption} \label{ass:Kass}
     For every $i \in [g], m \in [M]$, $f^{(i)} (\cdot,m)$ is a local contraction, that is, there exists constants $ L_{m}^{(i)} \in [0,1) $, $ c_{m}^{(i)} > 0 $ such that
    \begin{equation} \label{eq:Kass}
        \| f^{(i)} (\theta_1,m) - f^{(i)} (\theta_2,m) \| \leq L_{m}^{(i)}  \| \theta_1 - \theta_2 \|
    \end{equation}
    for every $\theta_1, \theta_2 \in \Theta^{(i)}$ with $\| \theta_1 - \theta_2 \| \leq c_{m}^{(i)} $.
\end{assumption} 
This assumption holds for the temperature dynamics in \eqref{eq:TCLdynamics}. 

\begin{figure}
    \centering
    \includegraphics[width=0.4\textwidth]{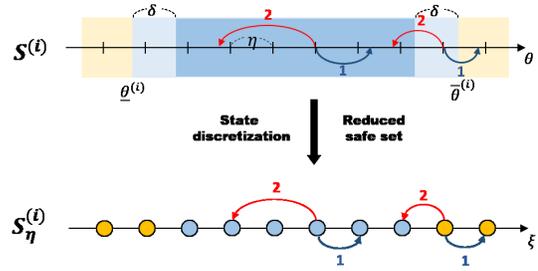}
    \vspace{-.1cm}
    \caption{Illustration of how the abstraction $S_{\eta}^{(i)}$ and its set of safe states $\Xi_{\text{safe}}^{(i)}$ are constructed. The state space $\Theta^{(i)}$ of subsystem $S^{(i)}$ is discretized to the set of grid points $\gamma_{\eta} (\Theta^{(i)})$. Each grid point transitions to the closest grid point to where it reaches in one time step, as the blue/red arrows indicate. The safe set $\Theta_{\text{safe}}^{(i)}$ (light blue) is reduced to $\Theta_{\text{safe}}^{(i)} \ominus \mathcal{B}(0,\delta)$ (darker blue) and the grid points intersecting with this set, which are the elements of $\gamma_{\eta} (\Theta_{\text{safe}}^{(i)} \ominus \mathcal{B}(0,\delta))$ (blue circles), are then the safe states of $S_{\eta}^{(i)}$.}
    \label{fig:abst_construction}
    \vspace{-0.5cm}
\end{figure}

 Now we construct an abstraction of $S^{(i)}$ for every $i \in [g]$ by uniformly discretizing $\Theta^{(i)}$. For a given grid size $\eta$, the abstraction function $\gamma_{\eta} : \mathbb{R}^d \to \mathbb{R}^d $ is defined as
\begin{equation}
    \gamma_{\eta}(\theta) = \eta \cdot \left \lfloor \frac{\theta}{\eta} \right \rfloor + \frac{\eta}{2} \bm{1}.
\end{equation}
 Suppose that the grid size for the abstraction corresponding to group $i$ is $\eta^{(i)}$, and define the set $\Xi^{(i)} := \gamma_{\eta} ( \Theta^{(i)} ) = \{ \widetilde{\xi}_{1}^{(i)},\ldots, \widetilde{\xi}_{K^{(i)}}^{(i)} \}$, where $K^{(i)}$ is the number of elements $\tilde{\xi}_k^{(i)}$. Then, an abstraction for $S^{ (i) }$ is 
    \begin{equation} \label{eq:dynoftsabst}
        S_{\eta^{(i)}}^{(i)} = ( \Xi^{(i)}, [M],\xrightarrow[(i),\eta^{(i)}]{}, \text{Id}_{\mathbb{R}^d}),
    \end{equation}
    where $\xi \xrightarrow[(i),\eta^{(i)}]{\mu} \xi' $ if and only if $\gamma_{\eta^{(i)}}(f^{ (i) } (\xi,\mu)) = \xi'$. This means that the transition relation between the states of $S_{\eta^{(i)}}^{(i)}$ is determined by propagating each grid point $\xi$ with the dynamics and finding the closest grid point $\xi'$ to where it reaches in one time step, as shown in Fig.~\ref{fig:abst_construction}. The abstraction of each subsystem $S_j^{(i)}$ is denoted $S_{j,\eta^{(i)}}^{(i)}$ and its state is denoted $\xi_{j}^{(i)}$. For simplicity, we drop the superscript $(i)$ from $\eta^{(i)}$ when not crucial.
    
    The following lemma, a discrete-time variant of those in \cite{pola2009symbolic,nilsson2016control}, states a condition for $S^{(i)}$ and $S_{\eta}^{(i)}$ to be bisimilar.
    \begin{lemma}\label{lemma:bisim_cond}
 
    If $\epsilon$ and $\eta$ are such that $(1-L_{m}^{(i)}) \epsilon \geq \eta/2$ for all $m \in [M]$,  then $S^{(i)}$ and $S_{\eta}^{(i)}$ are $\epsilon-$approximately bisimilar.  
    \end{lemma} 
    
    If $S^{(i)}$ and $S_{\eta}^{(i)}$ are $\epsilon$-approximately bisimilar, the states of $S^{(i)}$ and $S_{\eta}^{(i)}$ remain $\epsilon$-close (i.e. $\| \theta (t) - \xi (t) \| \leq \epsilon$) if the same input sequence is applied to both systems from the initial state $\theta(0)$, which satisfies $\| \theta(0) - \xi(0) \| \leq \epsilon$. 
    
    By Assumption~\ref{ass:Kass}, we can show that, for any $0 < \epsilon \leq \min_{ m \in [M]} \kappa_{m}^{(i)}$, there is a small enough $\eta>0$ that satisfies the condition in Lemma~\ref{lemma:bisim_cond}. Moreover, even when small disturbances or parameter variations exist among the subsystems in the same group, a single $\epsilon$-approximately bisimilar system for the entire group can be constructed. In particular, for any $m$ in $[M]$, let $\overline{\delta}_m$ be a constant which satisfies $(1-L_m^{(i)}) \epsilon \geq \eta^{(i)} / 2 + \overline{\delta}_{m} $. Then,  any subsystem whose state evolution function $f'$ satisfies
    \begin{equation*}
        \| f' (\theta,m) - f^{(i)} (\theta,m) \| \leq \overline{\delta}_m \quad \forall \theta \in \Theta^{(i)}, m \in [M],
    \end{equation*} 
    is also $\epsilon$-approximately bisimilar to $S_{\eta}^{(i)}$. This allows for mild heterogeneity within a group as long as every subsystem in it satisfies the inequality above.

    The abstraction of each group $i$ can be represented by the product transition system $S_{\eta}^{(i) \times N^{(i)}}$ and the abstraction of the entire system can be represented as the product of those systems, denoted $\bm{S}_{\bm{\eta}} := S_{\eta^{(1)}}^{(1) \times N^{(1)}} \times \cdots \times S_{\eta^{(g)}}^{(g) \times N^{(g)}}$, where $\bm{\eta} := (\eta^{(1)},\ldots,\eta^{(g)})^\top$. By definition~\ref{def:epsbisimilar}, it is easy to see that $\bm{S}$ and $\bm{S}_{\bm{\eta}}$ are $\epsilon$-approximately bisimilar if $S^{(i)}$ and $S_{\eta^{(i)}}^{(i)}$ are $\epsilon$-approximately bisimilar for all $i \in [ g ] $. Then, we can show the following.
\begin{theorem} \label{thm:relmaxinv1}
        Assume that $S^{(i)}$ and $S_{\eta}^{(i)}$ are $\epsilon$-approximately bisimilar and $ \delta$ is larger than $\epsilon + \eta^{(i)} / 2 $. If the abstracted subsystem $ S_{j,\eta}^{(i)}$ starting at $ \xi_j^{(i)} (0) $ stays inside $\gamma_{\eta^{(i)}} (\Theta_{\text{safe}}^{(i)} \ominus \mathcal{B}(0,\delta))$ under an input trajectory $\mu_j^{(i)} (t)$, then for every $\theta_j^{(i)} (0) \in \Theta^{(i)}$ satisfying $\| \theta_j^{(i)} (0) - \xi_j^{(i)} (0) \| \leq \epsilon$, the trajectory of $S_j^{(i)} $ starting at $\theta_j^{(i)} (0)$ stays inside the safe set $\Theta_{\text{safe}}^{(i)  }$ under $\mu_j^{(i)}(t)$.
\end{theorem}
This theorem states that, if we can obtain an input sequence for an abstracted subsystem that keeps its state inside a safe set shrunken by $\delta > \epsilon + \eta^{(i)} / 2$, then the input sequence is also safe for the original subsystem.

Using Theorem~\ref{thm:relmaxinv1}, we define the safe set for abstracted subsystem as $\Xi_{\text{safe}}^{(i)} := \gamma_{\eta^{(i)}}(\Theta_{\text{safe}}^{(i)} \ominus \mathcal{B}(0,\delta))$ and impose the following local constraints on the abstracted subsystems.
\begin{equation} \label{eq:abstlocalcond}
    \xi_j^{(i)} (t) \in \Xi_{\text{safe}}^{(i)}  \quad \forall i \in [g], \enspace j \in [N^{(i)}], \enspace t \in \mathbb{N}_0.
\end{equation}
Fig.~\ref{fig:abst_construction} illustrates how $S_{\eta}^{(i)}$ and its set of safe states $\Xi_{\text{safe}}^{(i)}$ are constructed.
Now our goal is to develop a safe control algorithm for the abstracted subsystems satisfying \eqref{eq:abstlocalcond}.

\subsection{Aggregate system}
We now construct an aggregate system for each group as an alternative representation of $S_{\eta}^{ (i) \times N^{(i)} }$. An aggregate system is a compact representation that keeps track of the number of subsystems in each possible combination of mode, state, and lockout duration (i.e., the number of time steps it has been locked in its current mode). The input to an aggregate system is the number of subsystems that switch mode, which can be used to determine the modes of the individual subsystems. To construct an aggregate system, we build a graph for each group that captures all possible transitions of the subsystem $S_{\eta}^{(i)}$ for all combinations of mode, state, and lockout duration. In particular, we extend the graph construction method from our preliminary work~\cite{jang2021large} by considering the lockout duration as one additional feature of a node. Then, the graph is used to express how the number of subsystems at each possible combination evolves, which acts as the dynamics of the aggregate system.
  
\begin{figure}
    \centering
    \includegraphics[width=0.45\textwidth]{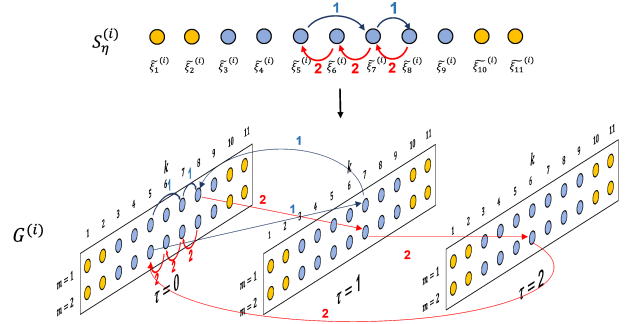}
    \vspace{-.2cm}
    \caption{The construction of an aggregate system graph $G^{(i)}$. The parameters are $M = 2$, $\overline{\tau}_1^{(i)} = 1$, $\overline{\tau}_2^{(i)} = 2$ in this example.
    } 
    \label{fig:graph_augmentation}
    \vspace{-0.5cm}
\end{figure}

We first construct a graph for the state-abstracted subsystem $ S_{\eta}^{(i)} $, i.e., $ G^{ (i) } = ( V^{ (i) } , E^{ (i) } ) $, where $V^{(i)}$ are the vertices and $E^{(i)}$ are the edges. Each node is denoted by $\nu_{m,\tau,k}^{ (i) }$ ($m \in [M]$, $\tau \in [\overline{\tau}_{m}^{(i)}]$, $k \in [ K^{(i)} ]$); when $S_{\eta}^{(i)}$ is at state $\widetilde{\xi}_{k}^{(i)}$, in mode $m$, and with lockout duration $\tau$, we say that $S_{\eta}^{(i)}$ `corresponds to' $\nu_{m,\tau,k}^{(i)}$. Edges are defined for any possible autonomous or controlled transition between nodes, specifically, $(\nu_{m_1,\tau_1,k_1}^{(i)},\nu_{m_2,\tau_2,k_2}^{(i)} )$ belongs to $E^{(i)}$ if $ \widetilde{\xi}_{k_1}^{(i)} \xrightarrow[(i),\eta]{m_2} \widetilde{\xi}_{k_2}^{(i)} $ and one of the following hold:
\begin{enumerate} \label{emu:graphtransitions}
    \item $ m_1 = m_2$, $\tau_1 = \tau_2 = 0 $ (system is unlocked and its mode remains unchanged),
    \item $ m_1 = m_2$, $\tau_1 > 0$, $\tau_2 = \tau_1 + 1 $ (system is locked and lockout duration is updated),
    \item
    $m_1 = m_2$, $\tau_1 = \overline{\tau}_{m_1}^{(i)}$, $\tau_2 = 0$ (system becomes unlocked),
    \item $m_1 \neq m_2$, $\tau_1 =0$,  $\tau_2 = \min (1,\overline{\tau}_{m_2}^{(i)})$ (system is unlocked and its mode changes).
\end{enumerate}
Fig.~\ref{fig:graph_augmentation} illustrates how $G^{(i)}$ is constructed.

Now, we define the \textit{safe nodes} of $G^{(i)}$ as the elements of the set
\begin{equation*}
    \begin{aligned}
    V_{\text{safe}}^{(i)} &:= \left \{ \nu_{m,\tau,k}^{(i)}  : \forall m \in [M], \tau \in [ \overline{\tau}_m^{(i)} ], k \in  \mathcal{I}_{\text{safe}}^{(i)} \right \},
    \end{aligned}
\end{equation*}
where $\mathcal{I}_{\text{safe}}^{(i)}$ is the set of indices $\{  k \in [ K^{(i)} ] : \widetilde{\xi}_{k}^{(i)} \in \Xi_{\text{safe}}^{(i)}\}$.
To abide by the local constraint for each subsystem~\eqref{eq:abstlocalcond}, every $S_{j,\eta}^{(i)}$ in group $i$ should correspond to an element of $V_{\text{safe}}^{(i)}$.

Then, the aggregate system is constructed with state $x^{(i)} \in \mathbb{N}_0^{D_x^{(i)} }$ (where $D_{x}^{(i)} := K^{(i)} \sum_{m=1}^M ( \overline{\tau}_{m}^{(i)}+ 1 )$), whose elements represent the number of subsystems in group $i$ corresponding to the nodes of $G^{(i)}$, and input $u^{(i)} \in \mathbb{N}_0^{D_{u}^{(i)}} $ (where $D_u^{(i)} := K^{(i)} M ( M-1)$), whose elements represent the number of subsystems switching mode from each node. Specifically, the element $x_{m,\tau,k}^{(i)}$ of $x^{(i)}$ represents the number of subsystems corresponding to $\nu_{m,\tau,k}^{(i)}$ and the element $u_{m_1,m_2,k}^{ (i) }$ of $u^{(i)}$ represents the number of unlocked subsystems corresponding to $\nu_{m_1,0,k}^{(i)}$ switching mode from $m_1$ to $m_2$. Then, the evolution of each element $x_{m,\tau,k}^{(i)} (t)$ is governed by
\begin{equation} \label{eq:elementevolution}
    x_{m,\tau,k}^{(i)}(t+1) = \begin{dcases}
        \sum_{\substack{\tau',k' : \\ (m,\tau',k') \in \mathcal{I}_{m,\tau,k}^{(i)}}} x_{m,\tau',k'}^{(i)} (t) & (\text{if } \tau \neq 1) \\
        \sum_{\substack{m',k' : \\
        (m',0,k') \in \mathcal{I}_{m,1,k}^{(i)}}} u_{m,m',k'}^{(i)} (t) & (\text{if } \tau = 1),
    \end{dcases}
\end{equation}
where the set $\mathcal{I}_{m,\tau,k}^{(i)}$ is the set of indices of the predecessors of $\nu_{m,\tau,k}^{(i)}$ defined as $\{ (m',\tau',k') :  (\nu_{m',\tau',k'},\nu_{m,\tau,k}) \in E^{(i)} \}$. 

Since every element of $x^{(i)}$ evolves as a linear combination of the state and input, the dynamics of $x^{(i)}$ can be written as the linear dynamics
\begin{equation} \label{eq:aggdynamics}
    \Gamma_{\eta}^{(i)} : x^{(i)} (t+1) = A^{(i)} x^{ (i) } (t) + B^{(i)} u^{(i)} (t),
\end{equation}
where $A^{(i)}$ and $B^{(i)}$ are determined by \eqref{eq:elementevolution} and depend upon the incidence matrices of $G^{(i)}$. Considering the number of subsystems $N^{(i)}$ in each group and the fact that only unlocked subsystems can switch mode, the state space $X^{(i)}$ and the admissible input space $U^{(i)} (x)$ of $\Gamma_{\eta}^{(i)}$ are as follows
\begin{equation} \label{eq:stateinputspace}
    \begin{aligned}
        & X^{(i)} =  \bigg \{ x^{(i)} \in  \mathbb{N}_0^{ D_x^{(i)}} : \sum_{m=1}^{M} \sum_{\tau=0}^{\overline{\tau}_{m}^{(i)}} \sum_{k=1}^{K^{(i)}} x_{m,\tau,k}^{(i)} = N^{(i)} \bigg \} \\
         & U^{(i)} (x^{(i)})  =  \Bigg \{ u^{(i)} \in \mathbb{N}_0^{D_u^{(i)}} : \\
          & \qquad \qquad \qquad \quad 0 \leq \sum_{ \substack{m_2 \in [M] \\  m_2 \neq m_1 }} u_{m_1,m_2,k}^{(i)} \leq x_{m_1,0,k}^{(i)} \Bigg \}.
    \end{aligned}
\end{equation}
Note that the lockout constraint~\eqref{eq:generallockout} is never violated if the number of subsystems switching mode at each node is an element in the admissible input set.

The local constraint imposed on $\Gamma_{\eta}^{(i)}$ is specified by the set $X_{\text{safe}}^{(i)}$, defined as follows
\begin{equation*}
    \begin{aligned}
        X_{\text{safe}}^{(i)} :=  \Big \{ x^{(i)} & \in X^{(i)} :  x_{m,\tau,k}^{(i)} = 0 \\ & \forall  m \in [M], \; \tau \in [ \overline{\tau}_{m}^{(i)}]_0, \; k \in \left ( [K^{(i)}] \setminus \mathcal{I}_{\text{safe}}^{(i)} \right ) \Big \},
    \end{aligned}
\end{equation*}
which requires all subsystems to correspond to safe nodes.

We can also construct an alternative representation of $\bm{S}_{\bm{\eta}}$ by combining all $\Gamma_{\eta}^{(i)}$, defined as $\bm{\Gamma}_{\bm{\eta}} := \Gamma_{\eta^{(1)}}^{(1)} \times \cdots \times \Gamma_{\eta^{(g)}}^{ ( g )}$. Accordingly, the state and input of $\bm{\Gamma}_{\bm{\eta}}$ are the concatenation of all $x^{(i)}$ and $u^{(i)}$, respectively, which are written as $\bm{x} := ( x^{(1) \top} , \ldots , x^{( g ) \top })^\top $, $\bm{u} : = ( u^{(1) \top}, \ldots, u^{( g ) \top} )^\top$. The state space and the admissible input space are $\bm{X} := X^{(1)} \times \cdots \times X^{ ( g )}$, $ \bm{U} ( \bm{x} ) := U^{(1)} (x^{(1)}) \times \cdots \times U^{ ( g )} ( x^{ ( g )} ) $, respectively. Then, the system dynamics of $\bm{\Gamma}_{\bm{\eta}}$ can be written as
\begin{equation} 
    \bm{\Gamma}_{\bm{\eta}} : \bm{x} (t+1) = \bm{A} \bm{x} (t) + \bm{B} \bm{u} (t),
\end{equation}
where the system matrices are $\bm{A} := \text{diag} ( \{ A^{ (i) } \}_{i=1}^{ g } ) $ and $\bm{B} : = \text{diag} ( \{ B^{ (i) } \}_{i=1}^{ g } )$. Hence, $\bm{\Gamma}_{\bm{\eta}}$ is a linear system with integer-valued state space and admissible input space. Crucially, the dimension of the state or input does not depend on the total number of subsystems $N^{(i)}$. 

Now, we define the set of safe states of $\bm{\Gamma}_{\bm{\eta}}$ comprising all states satisfying the local and global constraints as follows
\begin{equation} \label{def:aggsafeset}
    \begin{aligned}
        \bm{X}_{\text{safe}} := \bigg \{ & \bm{x}  \in \bm{X} : x^{(i)} \in X_{\text{safe}}^{(i)}  \enspace \forall i \in [ g ], \\
         \underline{P}_{m} \leq &  \sum_{i=1}^{g}  p_{m}^{ (i) } \sum_{\tau=0}^{\overline{\tau}_{m}^{(i)}} \sum_{k=1}^{K^{(i)}} x_{m,\tau,k}^{(i)} \leq \overline{P}_m \enspace \forall m \in [M] \bigg \},
    \end{aligned}
\end{equation}
where the first line enforces the local constraints on all subsystems, and the second line enforces the global constraint~\eqref{eq:generalaggconstraints}; note that $\sum_{\tau=0}^{\overline{\tau}^{(i)}_{m}} \sum_{k=1}^{K^{(i)}} x_{m,\tau,k}^{(i)}$ equals the number of the subsystems in mode $m$. All constraints within \eqref{def:aggsafeset} are linear inequalities.

Our objective is to develop a control algorithm for $\bm{\Gamma}_{\bm{\eta}}$ that keeps the system safe. Our algorithm, which will be introduced in Section~\ref{sec:INVMPC}, incorporates a controlled invariant set to guarantee recursive safety and feasibility. However, the large dimension of $\bm{\Gamma}_{\bm{\eta}}$ makes it difficult to use existing invariant set computation tools for linear systems \cite{nilsson2019control,anevlavix2019computing,wintenberg2020implicit}. Therefore, in the next section, we propose a method that obtains an implicit representation of an invariant set of $\bm{\Gamma}_{\bm{\eta}}$, which can be leveraged in the control algorithm.

\section{Construction of an invariant set by an implicit representation} \label{sec:invsetmethod}
In this section, we show how to construct a controlled invariant set for the aggregate system $\bm{\Gamma}_{\bm{\eta}}$. Leveraging the cycles of the graphs, this method finds an implicit representation of an invariant set of $\bm{\Gamma}_{\bm{\eta}}$ by identifying a set of states that can be kept inside $\bm{X}_{\text{safe}}$ via periodic inputs. This method is an extension of the method in our preliminary work~\cite{jang2021large} to accommodate aggregate systems with more than one group.
  \begin{figure}
      \centering
      \includegraphics[width=0.45\textwidth]{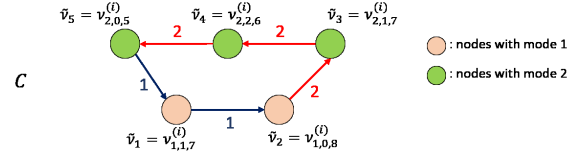}
      \caption{From the graph $G^{(i)}$ in Fig.~\ref{fig:graph_augmentation}, a safe cycle $C$ can be found with node sequence $(\nu_{1,1,7}^{(i)},\nu_{1,0,8}^{(i)},\nu_{2,1,7}^{(i)},\nu_{2,2,6}^{(i)},\nu_{2,0,5}^{(i)})$ and corresponding mode sequence $(1,1,2,2,2)$.} 
      \label{fig:CycleExample}
      \vspace{-0.5cm}
  \end{figure}

We first define a \textit{safe cycle} as a sequence of nodes $ \tilde{\nu} = (\tilde{\nu}_1,\ldots,\tilde{\nu}_{|C|}) \in (V_{\text{safe}}^{(i)})^{|C|}$ with corresponding mode sequence $\tilde{\mu} = (\tilde{\mu}_1,\ldots,\tilde{\mu}_{|C|}) \in [M]^{|C|}$ that satisfies $\tilde{\nu}_{l} \xrightarrow[(i),\eta]{\tilde{\mu}_{l+1}} \tilde{\nu}_{l+1}$ if $l \neq |C|$ and $\tilde{\nu}_{|C|} \xrightarrow[(i),\eta]{\tilde{\mu}_{1}} \tilde{\nu}_{1}$ if $l = |C|$. Since every node of a safe cycle belongs to $V_{\text{safe}}^{(i)}$, if an abstracted subsystem $S_{j,\eta}^{(i)}$ corresponds to one of these nodes at time $t$, then the subsystem satisfies the local constraint \eqref{eq:abstlocalcond} at time $t$. Note that at least one safe cycle should exist in each $G^{(i)}$ for the existence of a controlled invariant set of $\bm{\Gamma}_{\bm{\eta}}$. Fig.~\ref{fig:CycleExample} shows an example of a safe cycle in $G^{(i)}$.
   
First, we introduce a strategy that uses a safe cycle to control each abstracted subsystem  without violation of its local constraint~\eqref{eq:abstlocalcond}. Consider an abstracted subsystem $S_{j,\eta}^{(i)}$ that corresponds to a node $\tilde{\nu}_l$ of the safe cycle $C$. If we apply the mode sequence $(\tilde{\mu}_{l},\ldots,\tilde{\mu}_{|C|},\tilde{\mu}_{1},\ldots,\tilde{\mu}_{l-1})$ periodically to this subsystem, it visits $(\tilde{\nu}_{l},\ldots,\tilde{\nu}_{|C|},\tilde{\nu}_{1},\ldots,\tilde{\nu}_{l-1})$ periodically so that it never leaves $V_{\text{safe}}^{(i)}$. This means that if $S_{j,\eta}^{(i)}$ corresponds to a node of a safe cycle, applying the mode sequence of the safe cycle periodically guarantees recursive satisfaction of \eqref{eq:abstlocalcond}. We call this a \textit{circular shift strategy}.
    
Next, we present a condition for recursive satisfaction of the global constraint \eqref{eq:generalaggconstraints} when every subsystem is controlled by a circular shift strategy. Note that \eqref{eq:generalaggconstraints} is a linear inequality with respect to the number of subsystems in each mode in each group. Moreover, the total number of subsystems in each mode in each group at each time is periodic since the mode of every subsystem $S_j^{(i)}$ is periodic under a circular shift strategy. This implies that, to ensure global constraint satisfaction, we only need to determine whether the maximum/minimum number of subsystems in each mode in each group within one period satisfies the global constraint. Further, these maximum/minimum values are functions of the nodes each subsystem corresponded to at the initial time step. Therefore, we can characterize an invariant set implicitly by a constraint on the initial number of subsystems corresponding to each node that guarantees the recursive satisfaction of the global constraint under a circular shift strategy.
   
Mathematically, we first select $n^{(i)}$ safe cycles from $G^{(i)}$ for all $i\in [g]$; at least one safe cycle should be selected from each group (i.e., $n^{(i)} > 0 $ for all $i$ in $[g]$). The cycles selected from graph $G^{(i)}$ are denoted $C_1^{(i)},\ldots,C_{ n^{(i)} }^{(i)}$ whose nodes are $\tilde{\nu}_{j}^{(i)} := (\tilde{\nu}_{j,1}^{(i)} ,\ldots, \tilde{\nu}_{j,l_{j}^{(i)}}^{(i)} )$ with corresponding mode sequence  $\tilde{\mu}_j^{(i)} := (\tilde{\mu}_{j,1}^{(i)}, \ldots,\tilde{\mu}_{j,l_j^{(i)}}^{(i)})$, where $l_j^{(i)}$ is the length of $C_{j}^{(i)}$. 

Next, we define the \textit{subsystem assignment} of $C_j^{(i)}$ as a vector $\beta_j^{(i)} \in \mathbb{N}_0^{l_j^{(i)}}$ whose $l$th element $[\beta_j^{(i)}]_{l}$ represents the number of subsystems corresponding to $\tilde{\nu}_{j,l}^{(i)}$. Since the state of every subsystem in group $i$ corresponds to only one node at a time, the following holds
   \begin{equation} \label{eq:sumofcycass}
     \sum_{j=1}^{n^{(i)}} \bm{1}^{\top} \beta_j^{(i)} = N^{(i)} \quad \forall i \in [ g ].    
   \end{equation}
Next, we consider how the number of subsystems corresponding to each node changes as the subsystems are controlled by a circular shift strategy. We first define the circulant matrix as
   \begin{equation*}
        \Psi_{l} := \begin{bmatrix}
            0 & 0 & \cdots & 0 & 1 \\
            1 & 0 &  \cdots & 0 & 0 \\
            0 & 1 & \cdots & 0 & 0 \\
            \vdots & \vdots & \ddots & \vdots & \vdots \\
            0 & 0 & \cdots & 1 & 0 
        \end{bmatrix} \in \mathbb{R}^{l \times l}.
    \end{equation*}
    Then, the number of subsystems at each node of $C_j^{(i)}$ after $q$-steps of circular shift from $\beta_j^{(i)}$ is $(\Psi_{l_j^{(i)}})^{q} \beta_j^{(i)}$.
   Also, the number of subsystems in mode $m$ corresponding to $C_j^{(i)}$ after $q$-steps of circular shift from $\beta_j^{(i)}$ is defined as
    \begin{equation} \label{eq:modecounts_beta}
        H_{m,q} ( \beta_j^{ (i) } ) := \sum_{l : \tilde{\mu}_{j,l}^{(i)}=m} \Big [ (\Psi_{l_j^{(i)}})^{q} \beta_j^{(i)} \Big ]_{l}.
    \end{equation}
     Note that $H_{m,q} (\beta_j^{(i)})$ is periodic with respect to $q$ with period $l_j^{(i)}$ because of the periodicity of $(\Psi_{l_j^{(i)}})^{q}$.
     
\begin{figure*} 
    \centering
    \includegraphics[width=0.8\textwidth]{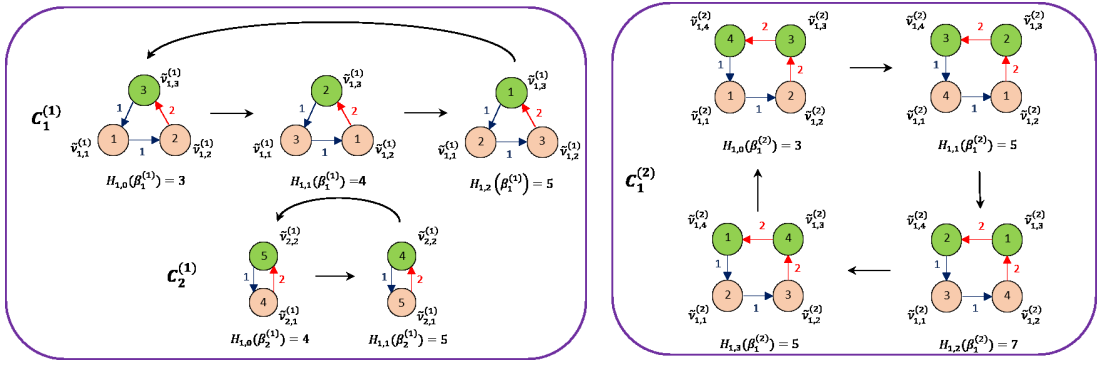}
    \vspace{-.1cm}
    \caption{In this example, we consider cycle assignments  $\beta_1^{(1)} = (1,2,3)$, $\beta_{2}^{(1)} = (4,5)$, $\beta_1^{(2)} = (1,2,3,4)$ with parameters $g = 2$, $n^{(1)} =2 $, $n^{(2)} = 1$, $p_{1}^{(1)} = p_{1}^{(2)} = 1$, and $\underline{P}_1 = 0, \overline{P}_1 = 19$. With $\max_{q' \in [3]} H_{1,q'} (\beta_1^{(1)}) = 5$, $\max_{q' \in [2]} H_{1,q'} (\beta_2^{(1)}) = 5$, and $\max_{q' \in [4]} H_{1,q'} (\beta_1^{(2)}) = 7$, the left side of \eqref{eq:safecycasscond2} becomes $5+5+7 = 17$. Since this is smaller than $\overline{P}_1$, \eqref{eq:safecycasscond2} holds for $m=1$. }
    \label{fig:cycle_ass}
     \vspace{-.5cm}
\end{figure*}

    Then, considering all selected cycles $C_1^{(1)},\ldots,C_{n^{(i)}}^{(i)}$ for group $i$, the number of subsystems in mode $m$ in group $i$ after $q$-steps of circular shift is $\sum_{j=1}^{ n^{(i)} } H_{m,q} (\beta_j^{(i)})$. Hence, the global constraint  \eqref{eq:generalaggconstraints} holds after $q$-steps of circular shift if the following holds
    \begin{equation} \label{eq:cycassaggconst}
            \underline{P}_m \leq \sum_{i=1}^{ g } p_m^{(i)} \sum_{j=1}^{n^{(i)}} H_{m,q} (\beta_j^{(i)}) \leq \overline{P}_m \quad \forall m \in [M].
    \end{equation}
    For recursive satisfaction of the global constraint under a circular shift strategy, we need \eqref{eq:cycassaggconst} to hold for all $q\in \mathbb{N}_0$. Considering the periodicity of each $H_{m,q}$ with respect to $q$, recursive satisfaction of \eqref{eq:generalaggconstraints} is achieved if the following two inequalities hold\footnote{These two inequalities are more conservative than the constraints introduced in our preliminary work~\cite{jang2021large}, yet, the number of inequalities needed is significantly less.}
    \begin{align} 
             &  \sum_{i=1}^{g }  p_{m}^{(i)} \sum_{j=1}^{ n^{(i)}}  \min_{q' \in [l_j^{(i)}]} H_{m,q'} (\beta_j^{(i)}) \geq \underline{P}_m  \quad \forall m \in [M], \label{eq:safecycasscond1} \\
             & \sum_{i=1}^{g } p_{m}^{(i)} \sum_{j=1}^{ n^{(i)}} \max_{q' \in [l_j^{(i)}]} H_{m,q'} (\beta_j^{(i)})  \leq \overline{P}_m \quad \forall m \in [M]. \label{eq:safecycasscond2} 
    \end{align}
    Fig.~\ref{fig:cycle_ass} illustrates one example of subsystem assignment for each cycle that satisfies those inequalities for $m=1$.
    
Now, a set $\Omega$ of subsystem assignments for which the circular shift strategy never violates the given constraints can be defined as
\begin{equation} \label{def:Omega}
        \begin{aligned}
             \Omega := \Bigg \{ ( \beta_1^{(1) \top},  \ldots,  & \beta_{n^{(g)}}^{( g ) \top} )^\top  \in  \mathbb{N}_0^{L}  : \enspace   \\ 
             & \text{s.t. } \eqref{eq:sumofcycass},\eqref{eq:safecycasscond1}, \eqref{eq:safecycasscond2} \text{ hold.} \Bigg \},
        \end{aligned}
\end{equation}
where $L$ is defined as $\sum_{i=1}^{g} \sum_{j=1}^{n^{(i)}} l_j^{(i)}$.

This implies that any state $\bm{x}$ of $\bm{\Gamma}_{\bm{\eta}}$ corresponding to a subsystem assignment in $\Omega$ will always stay inside $\bm{X}_{\text{safe}}$ using a circular shift strategy. To obtain a set $\bm{X}_{\text{inv}}$ of such $\bm{x}$, we use a mapping $\Phi_j^{(i)}$ from subsystem assignment vectors $\beta_{j}^{(i)}$ to the state of $\Gamma_{\eta^{(i)}}^{(i)}$. Specifically, if $x^{(i)} =  \sum_{j=1}^{n^{(i)}} \Phi_j^{(i)} ( \beta_j^{(i)})$, its element $x_{m,\tau,k}^{(i)}$ is 
\begin{equation*}
    x_{m,\tau,k}^{(i)} = \sum_{j=1}^{n^{(i)}} \sum_{l: \tilde{\nu}_{j,l}^{(i)} = \nu_{m,\tau,k}^{(i)}}  [ \beta_j^{(i)} ]_{l} .
\end{equation*}
Then using this mapping, we can define the projection of $\Omega$ into the state space $\bm{X}$ of $\bm{\Gamma}_{\bm{\eta}}$ as 
\begin{equation} \label{def:invset}
    \begin{aligned}
        \bm{X}_{\text{inv}} := \Bigg \{ \bm{x} \in \bm{X} : \enspace  \exists (\beta_1^{(1) \top},  \ldots, \beta_{n^{(g)}}^{ ( g ) \top})^\top \in \Omega, \\
         \text{s.t. }  x^{(i)}  = \sum_{j=1}^{n^{(i)}} \Phi_{j}^{(i)} (\beta_j^{(i)}) \enspace \forall i \in [ g ]\Bigg \}.
    \end{aligned}
\end{equation}
Therefore, the following theorem holds.
\begin{theorem}\label{theo:Xcyc_inv}
    $\bm{X}_{\text{inv}}$ is a controlled invariant set of the system $\bm{\Gamma}_{\bm{\eta}}$ under the constraint set $\bm{X}_{\text{safe}}$.
\end{theorem}

Note that the definition \eqref{def:invset} is not an explicit representation of $\bm{X}_{\text{inv}}$ since we use additional variables $(\beta_1^{(1)},\ldots,\beta_{n^{(g)}}^{(g)})$ in its description. To obtain an explicit representation that only depends on constraints in $x^{(i)}$, one needs to project the implicit representation \eqref{def:invset} to the state space $\bm{X}$. However, because of the large dimension of the set, the computational burden is large and thus the projection is usually intractable~\cite{tiwaryHardnessComputingIntersection2008}. Fortunately, to check whether a state $\bm{x}$ belongs to $\bm{X}_{\text{inv}}$ or to compute an input $\bm{u} \in \bm{U} (\bm{x})$ that guarantees invariance, an explicit representation is not necessary. Instead, if we introduce integer variables encoding the minimum/maximum in \eqref{eq:safecycasscond1}-\eqref{eq:safecycasscond2}, $\bm{X}_{\text{inv}}$ can be represented with linear inequalities in $x^{(i)}$, $\beta_j^{(i)}$, and these integer variables, and these inequalities can be used in our formulation.
    
\section{Implicit invariant-set-driven MPC} \label{sec:INVMPC}

    In this section, we propose an MPC-based control algorithm with recursive safety and feasibility, referred to as \textit{Implicit Invariant-Set-Driven MPC}.
    Suppose the state of $\bm{\Gamma}_{\bm{\eta}}$ is $\bm{x} (t)$ at time step $t$. Then, this algorithm solves the following program at every time step $t \in \mathbb{N}_0$ for a given horizon length $h$,
    \begin{subequations}  \label{alg:MPCwithimplicit} 
    \begin{align}
        \min & \enspace \sum_{\tau=0}^{h} \hat{c}_{t+\tau} (\bm{x}^{\tau|t}) \nonumber \\
        \text{s.t.} 
        & \enspace \bm{x}^{\tau+1|t} = \bm{A} \bm{x}^{\tau|t} + \bm{B} \bm{u}^{\tau|t}  & \forall \tau \in [h-1]_0 \label{eq:INVMPCconst1} \\ 
        & \enspace \bm{u}^{\tau|t} \in \bm{U} ( \bm{x}^{\tau|t} )  & \forall \tau \in [h-1]_0 \label{eq:INVMPCconst2} \\
        & \enspace \bm{x}^{0|t} = \bm{x}(t) \label{eq:INVMPCconst3} \\
        & \enspace (\bm{x}^{h|t})^{(i)} =  \sum_{j=1}^{n^{(i)}} \Phi_{j}^{(i)} (\beta_j^{(i)}) \label{eq:INVMPCconst4} \\
        & \enspace (\beta_{1}^{(1) \top},\ldots,\beta_{n^{(g)}}^{(g) \top})^\top \in \Omega. \label{eq:INVMPCconst5}
    \end{align}
    \end{subequations}
    The decision variables are $\bm{x}^{\tau|t} \in \mathbb{N}_0^{ \sum_i D_x^{(i)}}$ for all $\tau$ in $[h]_0$, $\bm{u}^{\tau|t} \in  \sum_{i} \mathbb{N}_0^{D_u^{(i)}} $ for all $\tau$ in $[h-1]_0$, and $\beta_j^{(i) \top} \in \mathbb{N}_0^{ l_j^{(i)}}$ for all $i \in [ g ]$, $j \in [n^{(i)}]$. The function $\hat{c}_{t}$ is the cost with respect to the state of $\bm{\Gamma}$ corresponding to $c_t$ in Problem~\ref{pr:generalproblem}.
    In \eqref{eq:INVMPCconst4}, $(\bm{x}^{h | t})^{(i)}$ is a vector with the entries of $\bm{x}^{h |t}$ corresponding to group $i$. After solving this problem, we select the optimal $\bm{u}^{0|t}$ as the input $\bm{u} (t)$ which drives the state to $ \bm{x} (t+1) = \bm{x}^{1|t}$. 
    
    To guarantee recursive safety and feasibility, this algorithm uses a similar scheme to those introduced in~\cite{koller2018learning,wabersich2018linear,wabersich2021probabilistic}, in which the state at the end of the horizon is required to belong to an invariant set. However, instead of using an explicit representation, our algorithm uses the obtained implicit representation to formulate the constraint~\eqref{eq:INVMPCconst4}. 

\begin{theorem}\label{thm:invsetMPCsafety}
    Suppose that the initial state $\bm{x} (0)$ belongs to the $h$-step backward reachable set of $\bm{X}_{\text{inv}}$. Then, the program \eqref{alg:MPCwithimplicit} has a feasible solution at any time step $t \in \mathbb{N}_0$. In addition, the trajectory $ \bm{x} (t)$ generated by the algorithm always belongs to the maximal controlled invariant set $\overline{\bm{X}}_{\text{inv}}$ of $\bm{\Gamma}_{\bm{\eta}}$, and therefore belongs to $\bm{X}_{\text{safe}}$, for every $t \in \mathbb{N}_0$.
\end{theorem}
Since the proof of this theorem is similar to the one of Theorems \RN{3}.5. and \RN{3}.7. of \cite{wabersich2018linear}, we do not provide it here.

 \begin{remark}
        The performance of implicit invariant-set-driven MPC in terms of cost minimization becomes better as the size of $\bm{X}_{\text{inv}}$ and the length of horizon $h$ increases, and the size of $\bm{X}_{\text{inv}}$ usually increases when the number of cycles $n^{(i)}$ selected from $G^{(i)}$ is larger. However, more cycles and larger $h$ mean more variables and constraints, requiring more computation time. Thus, $h$ and $n^{(i)}$ should be carefully chosen considering this trade-off.
 \end{remark}
 
 To implement the control input, we arbitrarily select $u_{m_1,m_2,k}^{(i)} (t)$ out of $x_{m_1,0,k}^{(i)} (t)$ unlocked subsystems in mode $m_1$ and state $k$ (i.e., subsystems corresponding to node $\nu_{m_1,0,k}^{(i)}$) and switch their mode to $m_2$. This process is repeated for all $m_2 \in [M] \setminus \{ m_1 \}$. The state of each subsystem at each time step can be determined from the sequence of states $\bm{x} (t)$ and inputs $\bm{u} (t)$ obtained by implicit invariant-set-driven MPC. 
    
By Theorems~\ref{thm:relmaxinv1} and~\ref{thm:invsetMPCsafety}, this switching control approach guarantees $\bm{S}$ always satisfies the constraints \eqref{eq:generalsafesetcond},\eqref{eq:generallockout},\eqref{eq:generalaggconstraints}. First, the lockout constraint~\eqref{eq:generallockout} is never violated since the controller only switches unlocked subsystems. Second, the global constraint~\eqref{eq:generalaggconstraints} holds by Theorem~\ref{thm:invsetMPCsafety}; the obtained state trajectory $\bm{x}(t)$ always belongs to $\bm{X}_{\text{safe}}$. Third and finally, the local constraint \eqref{eq:generalsafesetcond} holds because every subsystem $S_j^{(i)}$ corresponds to one of the nodes in $V_{\text{safe}}^{(i)}$, i.e, state $\xi_j^{(i)} (t) $ always belongs to $\gamma_{\eta^{(i)}} ( \Theta_{\text{safe}}^{(i)} \ominus \mathcal{B} ( 0,\delta))$, and so, by Theorem~\ref{thm:relmaxinv1}, the state trajectory $\theta_j^{(i)} (t)$ of each $S_j^{(i)}$ always belongs to $\Theta_{\text{safe}}^{(i)}$.

\section{Simulation Results} \label{sec:simulation}

 In this section, we compare the performance of implicit invariant-set-driven MPC to several benchmark algorithms by solving Problem~\ref{pr:TCLproblem}. We assume that two groups of homogeneous TCLs (i.e. $g=2$) are connected to the 56-bus balanced distribution network in~\cite{bolognani2015existence}. \footnote{The parameters used are as follows: $T_a^{(1)} = T_{a}^{(2)} = 32$ \textdegree{}C, $C_{\text{th}}^{(1)} = 1.8$~kWh/\textdegree{}C, $C_{\text{th}}^{(2)} = 2.0$ kWh/\textdegree{}C, $R_{\text{th}}^{(1)} = 1.5$ \textdegree{}C/kW, $ R_{\text{th}}^{(2)} = 2.0$ \textdegree{}C/kW, $p_{\text{tr}}^{(1)} = 16$ kW, $p_{\text{tr}}^{(2)} = 14$ kW, $\underline{T}^{(1)} = 21.25$ \textdegree{}C, $\underline{T}^{(2)} = 23.25$ \textdegree{}C, $\overline{T}^{(1)} = 23.75$ \textdegree{}C, $\overline{T}^{(2)} = 25.75$ \textdegree{}C,  $N_{\text{TCL}}^{(1)}=250$, $N_{\text{TCL}}^{(2)}=210$,  $\overline{t}_{\text{on}}^{(1)} = \overline{t}_{\text{on}}^{(2)} = 150$s,  $\overline{t}_{\text{off}}^{(1)} = \overline{t}_{\text{off}}^{(2)} = 30$s.} The uncontrollable load at each node is assumed to be constant with the value of 50\% of the nominal power consumption at each node; the other 50\% is the nominal power consumption of controllable TCLs. We use a time discretization $\Delta t$ of the TCL dynamics of $40$s.
    
For the safety of the network, we try to prevent under-voltages, where the lower bound on voltage is set to 0.95 p.u.. From this bound on voltage, the safe upper bound on aggregate power $\overline{P}_{\text{agg}}$ is obtained using the method proposed in \cite{ross2020method}.  Note that our preliminary work~\cite{jang2021large} did not explicitly consider a network model, or show the impact of TCL control on the network, like we do here.
    
The reference signal $r(t)$ for aggregate TCLs is obtained by shifting and scaling a PJM frequency regulation signal~\cite{PJMrefsignal} from 13:00 to 14:30. We both shift and scale it by the nominal power consumption of the controllable TCLs. The cost function to measure the tracking performance is 
    \begin{equation}
        \hat{c}_t ( \bm{x} ) =  \left | \sum_{i=1}^{g}  p^{(i)} \sum_{ \tau=0 }^{\overline{\tau}_{\text{on}}^{(i)}} \sum_{k=1}^{K^{(i)}} x_{1,\tau,k}^{(i)} - r (t) \right |.
    \end{equation}
    The number of cycles selected for implicit invariant-set-driven MPC for each group is $n^{(1)} = n^{(2)} = 8$. The parameters for the abstraction are  $\epsilon = 0.8$, $\eta^{(1)} = 0.0066$, $\eta^{(2)} = 0.0055 $, and $\delta = 0.8033$. Gurobi is used to solve the optimization problems, and the limitation on the solve time of each iteration of MPC is set to 1000s; if the solver cannot find an optimal solution in 1000s, a suboptimal solution is used.

\subsection{Benchmark algorithms} \label{subsec:Benchmark}
     Three benchmark algorithms are compared with implicit invariant-set-driven MPC.
    \subsubsection{Benchmark 1} \label{alg:Benchmark1}
    This MPC algorithm requires the states to be inside $\bm{X}_{\text{safe}}$ over the optimization horizon. The problem solved at time $t$ is 
    \begin{equation*} 
        \begin{aligned}
            \min & \enspace \sum_{\tau=0}^{h} \hat{c} (\bm{x}^{ \tau |t}, r(t+\tau))  \\
            \text{s.t.} 
            & \enspace \eqref{eq:INVMPCconst1},\eqref{eq:INVMPCconst2},\eqref{eq:INVMPCconst3}  \\
            & \enspace \bm{x}^{\tau|t} \in \bm{X}_{\text{safe}} \quad \quad \quad \quad  \forall \tau \in [h]_0. \\
        \end{aligned}
    \end{equation*}
    When this problem is feasible, the obtained solution keeps the state inside $\bm{X}_{\text{safe}}$. However, this algorithm does not ensure recursive feasibility. Therefore, the state may go outside of the maximal controlled invariant set and the problem may become infeasible.

    \subsubsection{Benchmark 2} \label{alg:Benchmark2}
    Instead of imposing a constraint for safety, this MPC algorithm tracks the reference signal truncated between the aggregate power limits (the blue line in Fig.~\ref{fig:trackingex}). The truncated reference signal is 
    \begin{equation*}
        \hat{r} (t) = 
        \begin{cases}
            r(t) & \text{if } \underline{P}_{\text{agg}} \leq r(t) \leq \overline{P}_{\text{agg}} \\ 
            \underline{P}_{\text{agg}} &  \text{if } r(t) < \underline{P}_{\text{agg}} \\
            \overline{P}_{\text{agg}} & \text{if } r(t) > \overline{P}_{\text{agg}}.
        \end{cases}
    \end{equation*}
    Then, the problem solved at time $t$ is
     \begin{align} 
            \min & \enspace \sum_{\tau=0}^{h} \hat{c} (\bm{x}^{\tau | t},\hat{r}(t+\tau)) \nonumber \\
            \text{s.t.}  & \enspace \eqref{eq:INVMPCconst1},\eqref{eq:INVMPCconst2},\eqref{eq:INVMPCconst3}.  \nonumber
    \end{align}
    Since this problem has fewer constraints, the computational burden of this algorithm is less than that of the others. However, solutions from this algorithm may violate the aggregate power limits if $\hat{r}$ is not perfectly tracked.
    
    \subsubsection{Benchmark 3} \label{alg:Benchmark3}
     The third benchmark is a version of invariant-set-driven MPC introduced in our preliminary work~\cite{jang2021large}, which does not consider lockout constraints.
     
     Note that Benchmarks 1 and 2 use the same aggregate system $\bm{\Gamma}_{\bm{\eta}}$ as implicit invariant-set-driven MPC, while Benchmark 3 leverages an aggregate system introduced in \cite{jang2021large}, whose state does not include locked durations. 

    \subsection{Numerical Experiments}

\subsubsection{Experiment 1}

\begin{figure}[t]
     \centering
     \begin{subfigure}[b]{0.23\textwidth}
         \centering
         \includegraphics[width=\textwidth]{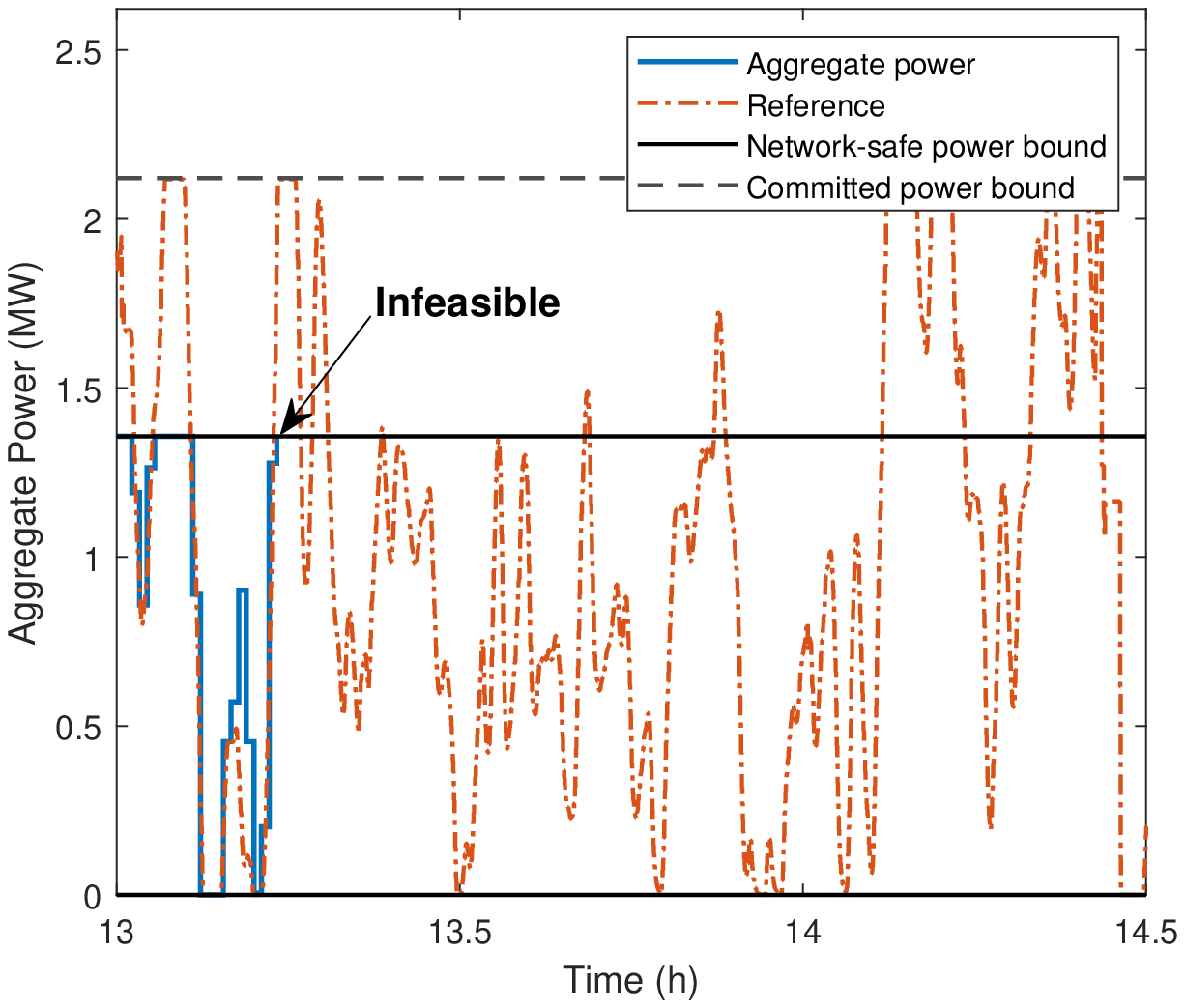}
         \caption{Benchmark 1}
         \label{fig:Benchmark1Exp1}
     \end{subfigure}
     \hfill
     \begin{subfigure}[b]{0.23\textwidth}
         \centering
         \includegraphics[width=\textwidth]{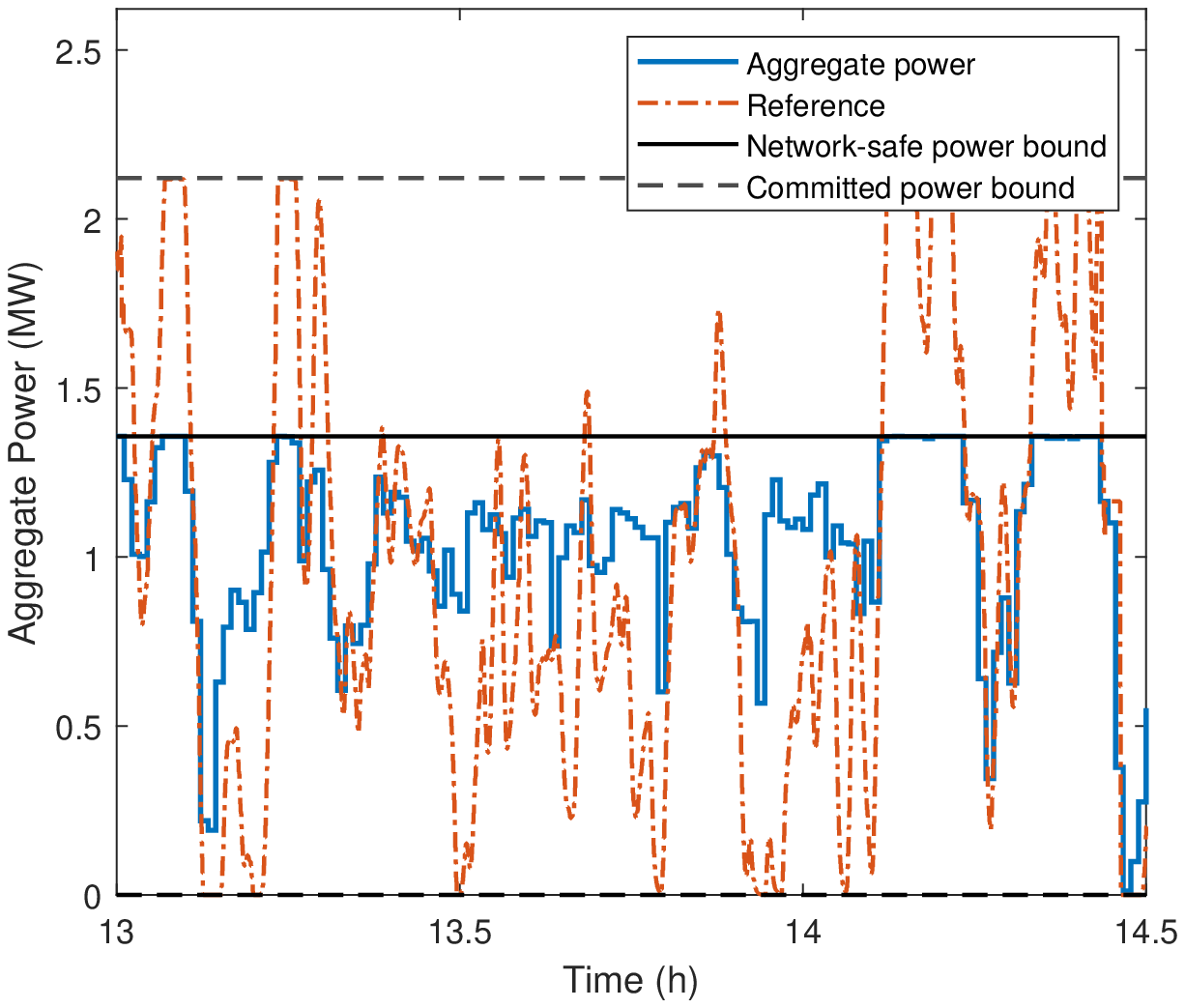}
         \caption{Invariant-set-driven MPC}
         \label{fig:INVMPCExp1}
     \end{subfigure}
    \caption{ Experiment 1: Since Benchmark 1 does not guarantee recursive feasibility, it becomes infeasible (left). In contrast, implicit invariant-set-driven MPC is recursively feasible and ensures safety at any time step (right).}
    \label{fig:Experiment1}
     \vspace{-.5cm}
\end{figure}

We first verify the recursive feasibility of implicit invariant-set-driven MPC in contrast to Benchmark~1. Fig.~\ref{fig:Experiment1} shows the results of Benchmark 1 and implicit invariant-set-driven MPC. Neither algorithm violates the bounds on aggregate power. Benchmark~1 shows better tracking performance at first. However, to achieve this, the temperatures of many of the TCLs in off-mode increase towards the upper bounds of their dead-bands. Eventually, no input exists that satisfies both the temperature and aggregate power constraints at the next time step. Thus, the problem becomes infeasible around 13.25h. In contrast, even though, initially, implicit invariant-set-driven MPC has worse tracking performance, it ensures feasibility at any time step by turning on some TCLs in advance.

\subsubsection{Experiment 2}

\begin{figure*}[!t]
         \centering
     \begin{subfigure}[b]{0.3\textwidth}
         \centering
         \includegraphics[width=\textwidth]{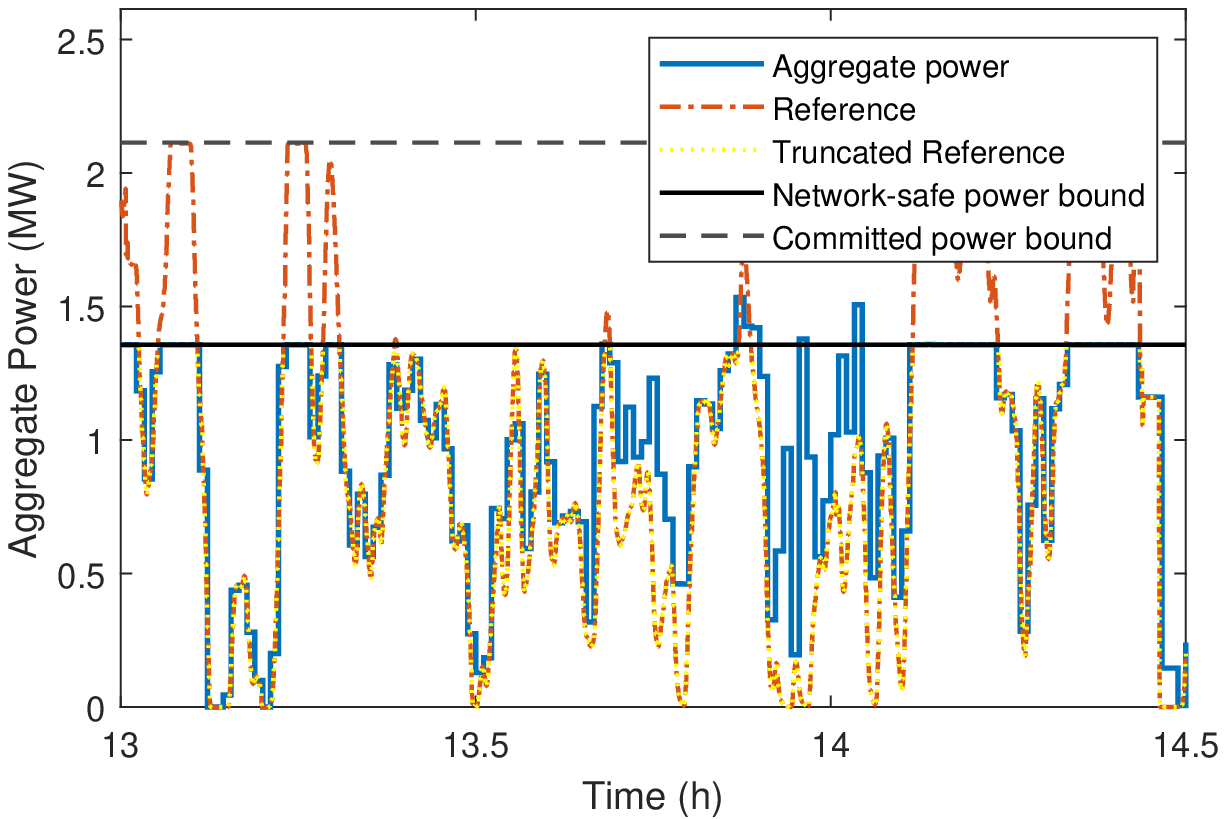}
         \label{fig:Benchmark2Exp2}
     \end{subfigure}
     \hfill
     \begin{subfigure}[b]{0.3\textwidth}
         \centering
         \includegraphics[width=\textwidth]{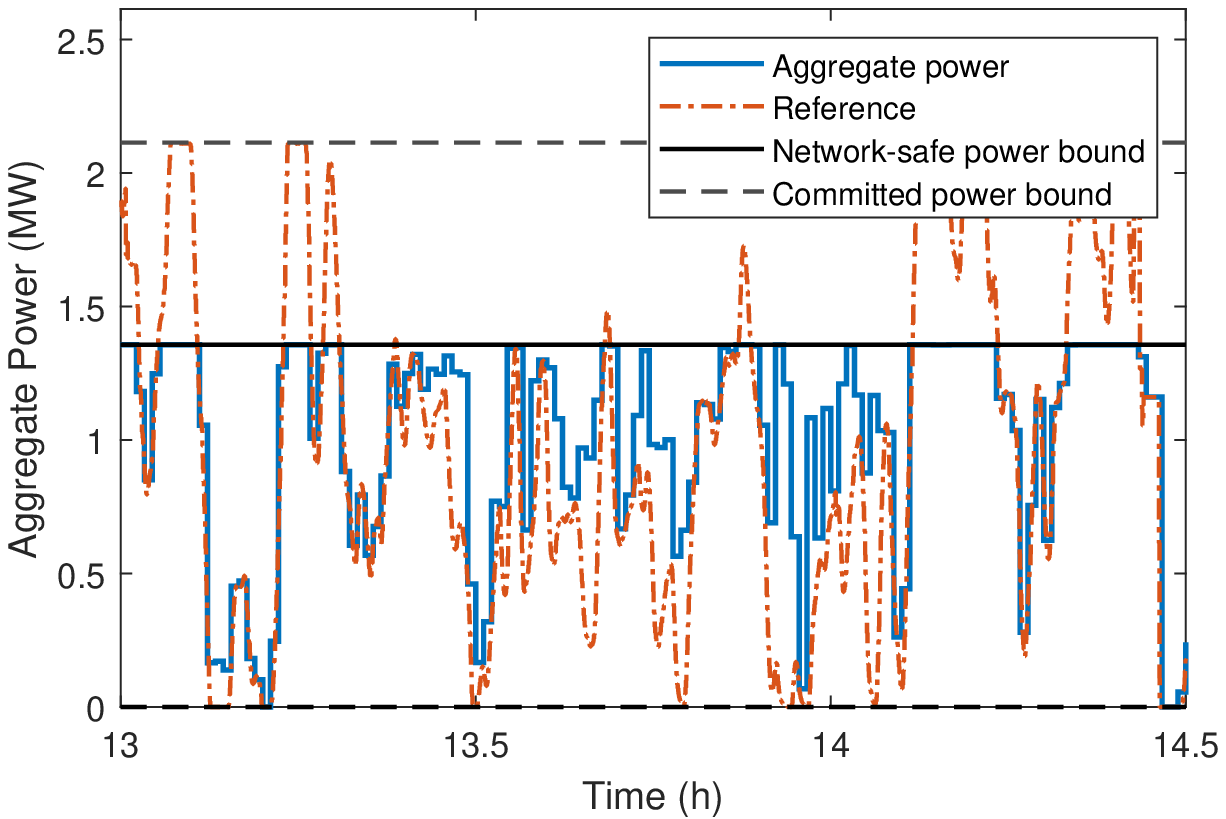}
         \label{fig:Benchmark3Exp2}
     \end{subfigure}
     \hfill
     \begin{subfigure}[b]{0.3\textwidth}
         \centering
         \includegraphics[width=\textwidth]{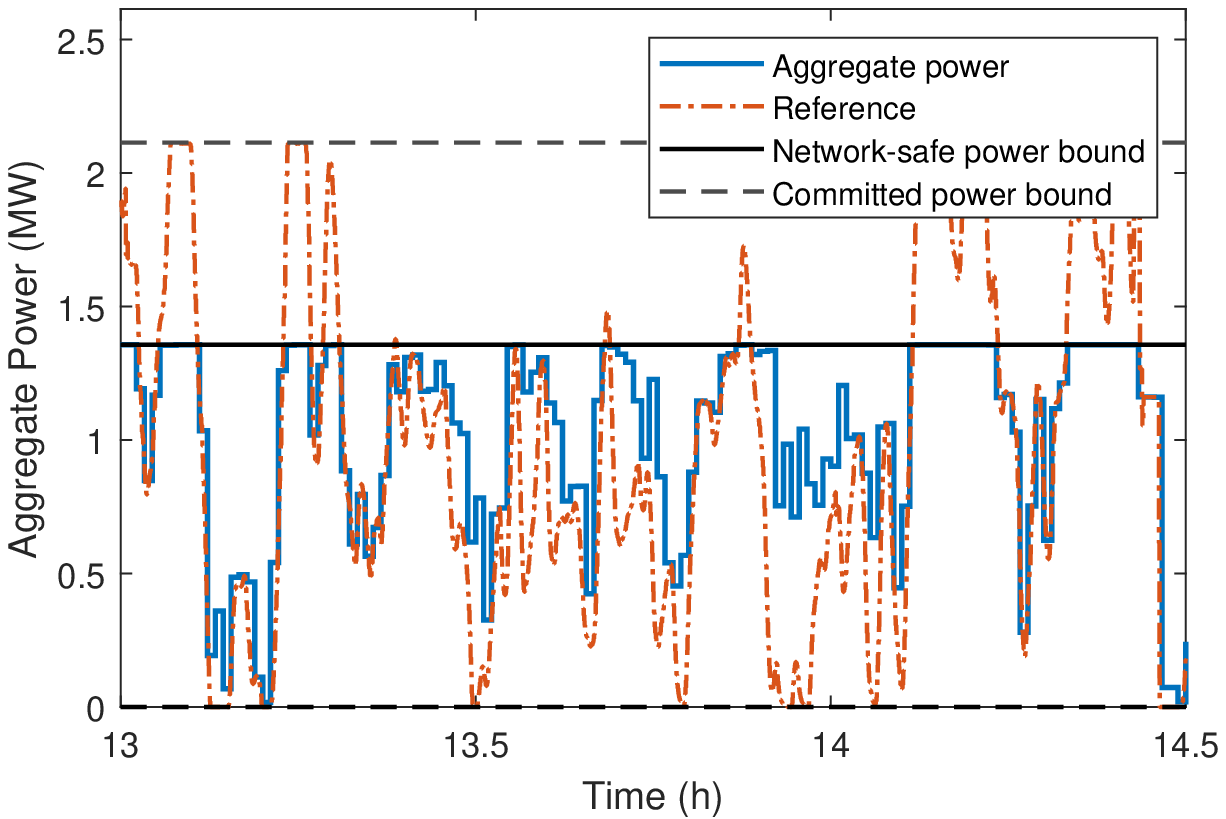}
         \label{fig:INVMPCExp2}
     \end{subfigure}
         \vspace{-.4cm}

    \begin{subfigure}[b]{0.3\textwidth}
         \centering
         \includegraphics[width=\textwidth]{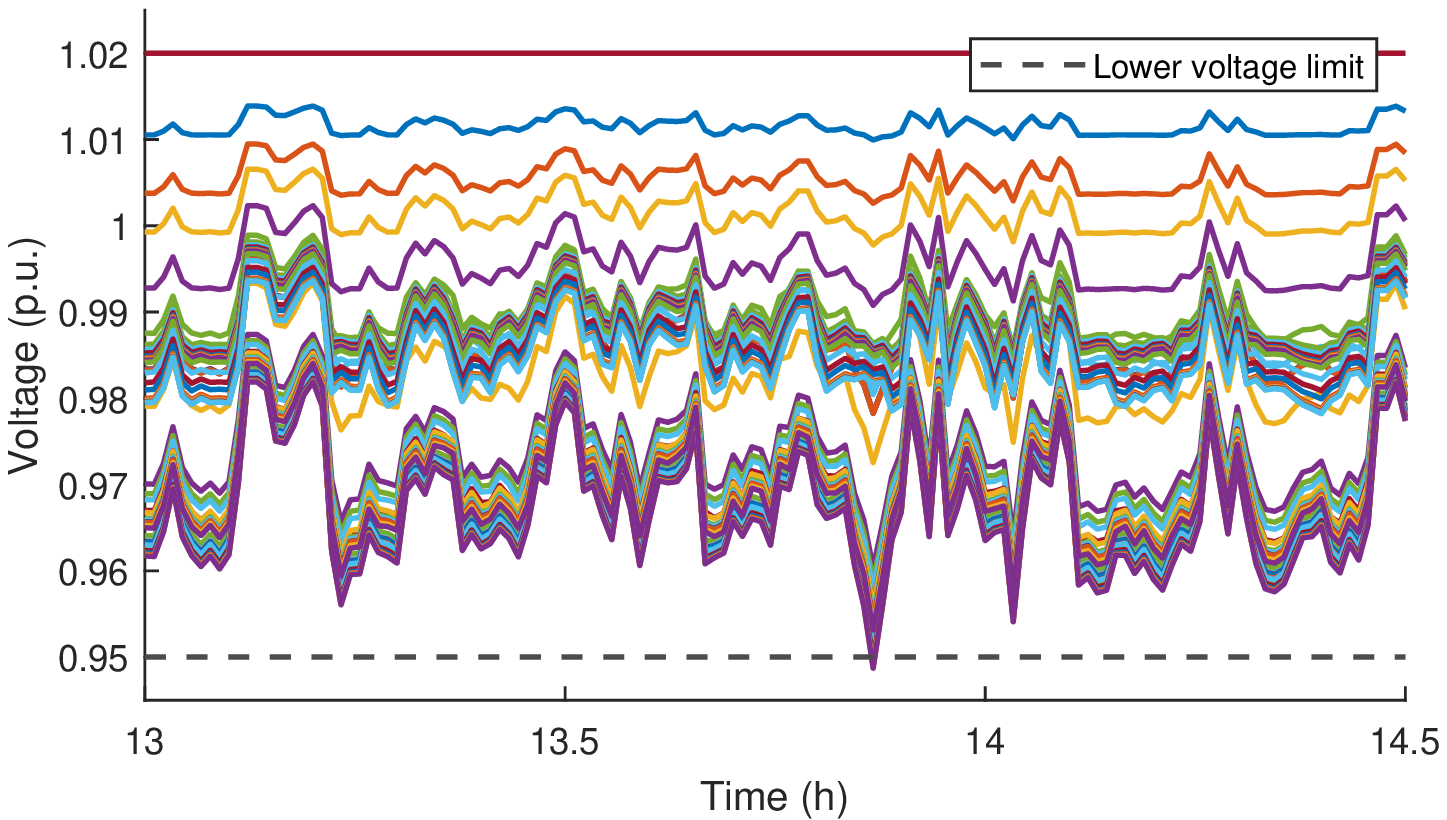}
          \vspace{-.4cm}
         \caption{Benchmark 2}
         \label{fig:Benchmark2Exp2Voltage}
     \end{subfigure}
     \hfill
     \begin{subfigure}[b]{0.3\textwidth}
         \centering
         \includegraphics[width=\textwidth]{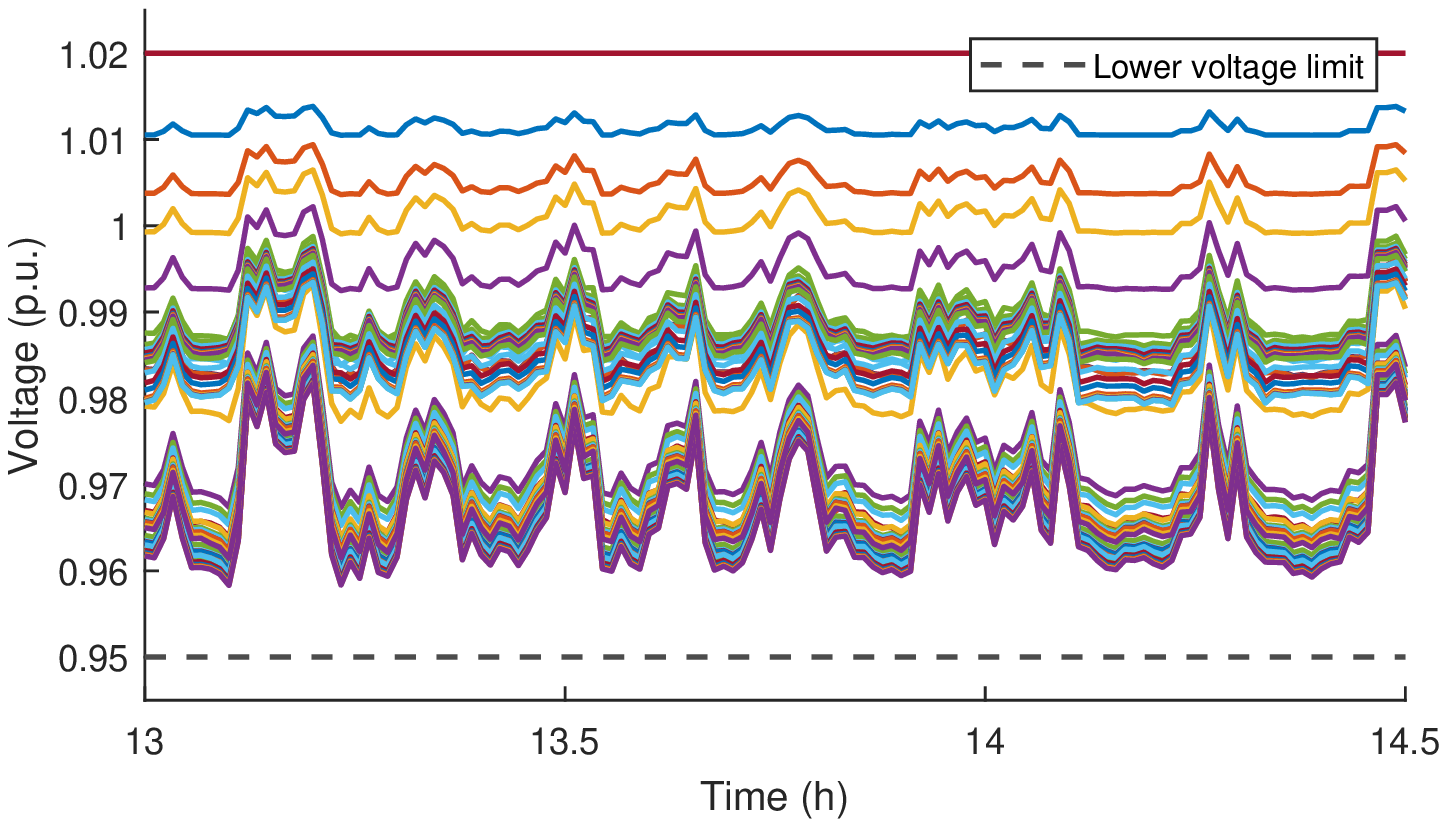}
          \vspace{-.4cm}
         \caption{Benchmark 3}
         \label{fig:Benchmark3Exp2Voltage}
     \end{subfigure}
     \hfill
     \begin{subfigure}[b]{0.3\textwidth}
         \centering
         \includegraphics[width=\textwidth]{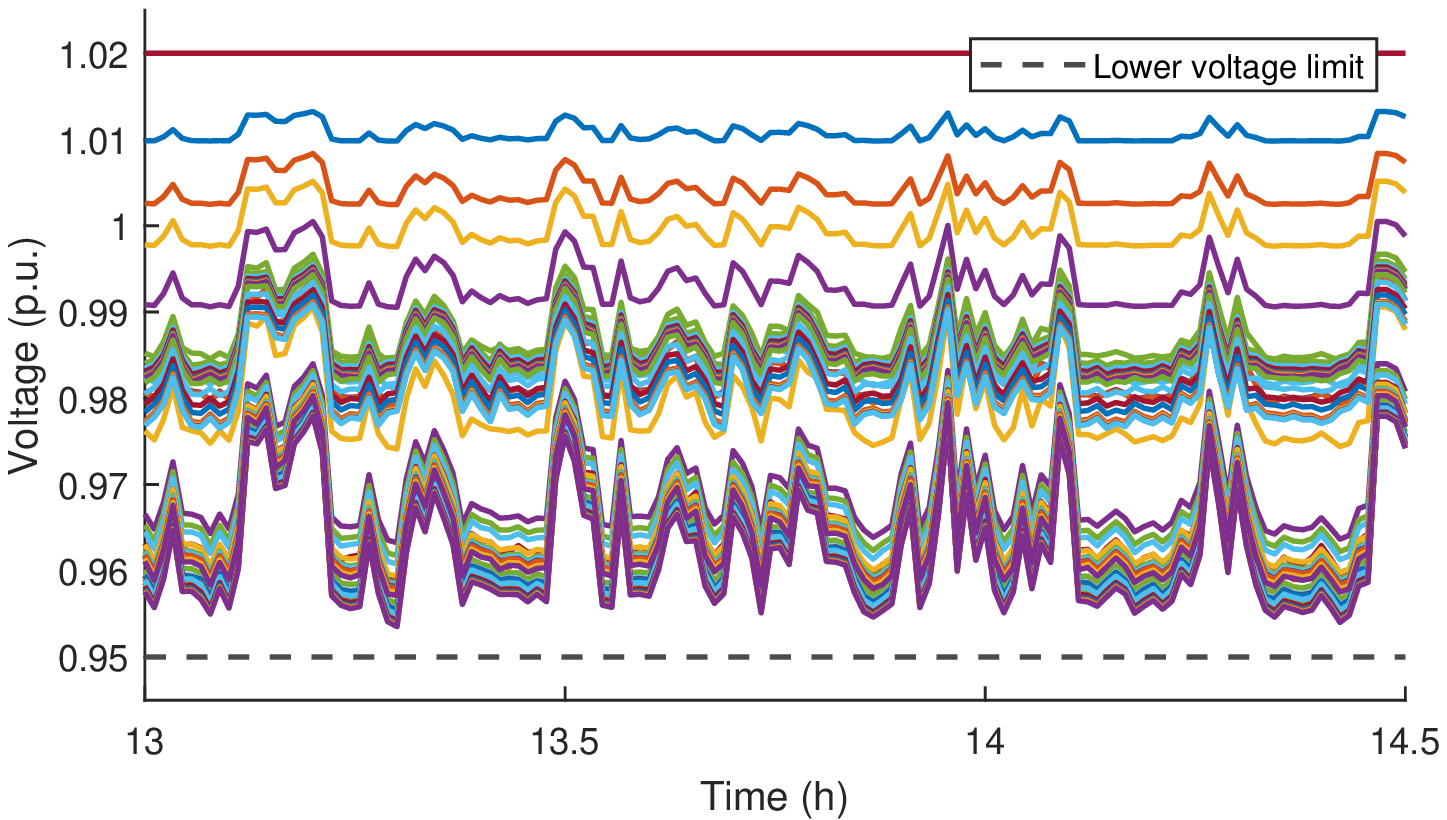}
         \vspace{-.4cm}
         \caption{Invariant-set-driven MPC}
         \label{fig:INVMPCExp2Voltage}
     \end{subfigure}
        \caption{Experiment 2: Benchmark 2 violates the network-safe power bounds and the lower voltage limit, while Benchmark 3 and Implicit invariant-set-driven MPC never do.}
        \label{fig:Experiment2}
         \vspace{-.5cm}
    \end{figure*}

In this experiment, we compare Benchmarks~2 and~3 with implicit invariant-set-driven MPC in terms of both tracking performance and constraint satisfaction. The results are shown in Fig.~\ref{fig:Experiment2} and Table~\ref{tab:Experiment2Perf}. Benchmark~2 has the best tracking performance, but the aggregate power exceeds the upper bound around 13.8h and 14.1h. Benchmark~3 never violates the bounds on aggregate power, but it makes an average of 12\% of TCLs violate their lockout constraints at each time step. In contrast, implicit invariant-set-driven MPC never violates the aggregate power or lockout constraints. 
    
The plots at the bottom of Fig.~\ref{fig:Experiment2} show the voltage at every node in the network for each algorithm. We see that Benchmark 2 leads to an under-voltage violation when the bound on aggregate power is violated. In contrast, Benchmark 3 and implicit invariant-set-driven MPC maintain safe voltages across the network. We note that the particular voltage violation seen in Fig.~\ref{fig:Experiment2} is small and unlikely to cause a real issue in the network, but other combinations of tracking signals, aggregate power bounds, and networks could lead to more severe violations that our approach would guard against.
    
The average time expended for each iteration of MPC is presented in Table~\ref{tab:Experiment2Perf}. Each iteration of implicit invariant-set-driven MPC takes much longer than that of Benchmark~2. Note that the average value exceeds the sampling time $\Delta t = 40s$. This means that the current implementation of the algorithm cannot be leveraged online and thus reducing the computation burden is an important topic for future work.
   
\begin{table}[t]
    \centering
    \caption{Performance comparison}
    \begin{tabular}{lccc}
\hline
& Benchmark 2 & Benchmark 3 & Invset MPC \\
\hline \hline
 RMSE (kW) & $3.95 \times 10^2$ 
  & $ 4.40 \times 10^2 $ & $4.37 \times 10^2$ \\
  \% of TCLs violating \eqref{eq:lockoutconstraints} \hspace{-.2cm}  & 0 & 12.42 & 0 \\
   Violation of \eqref{eq:TCLaggconstraint} & Yes & None & None \\
 Average Time (s) & $2.18$ & $131.46 $ & $73.37$ \\   
  \hline
\end{tabular}
    \vspace{-.3cm}
    \label{tab:Experiment2Perf}
\end{table}  
    
\section{Conclusions} \label{sec:conclusion}

In this paper, we proposed an invariant set construction method that provides an implicit representation of the safe set of a system of heterogeneous switched subsystems with both local and global safety constraints. This representation is incorporated, via a terminal condition, into our proposed MPC-based control algorithm. Since the computational burden of this algorithm is independent of the number of subsystems, it is scalable to large collections of subsystems. Numerical simulation results demonstrated the safety and recursive feasibility of the approach.
    
Possible future work includes the development of data-driven abstractions to allow for more heterogeneity. From an application-domain standpoint, it would also be valuable to develop approaches that handle time-varying power bounds, temperature setpoints, ambient temperatures, and disturbances.

\bibliographystyle{IEEEtran}
\bibliography{reference} 

\section*{Appendix}

\subsection*{Supplements \& Proofs for Section~\ref{sec:abstraction} :}

\begin{proof} [Lemma~\ref{lemma:bisim_cond}]
It follows from \cite{nilsson2016control} where $\alpha_m(x) = L_m^{(i)} x$ acts as the class $\mathcal{K}$-function in their proof for a fixed sampling time.
\end{proof}
The following lemma is used to prove Theorem~\ref{thm:relmaxinv1}.

\begin{lemma} \label{lem:invlemma1}
    If $\delta > \epsilon + \eta^{(i)} /2$ holds, $\gamma_{\eta^{(i)}} (\Theta_{\text{safe}}^{(i)} \ominus \mathcal{B}(0,\delta))$ is a subset of $\Theta_{\text{safe}}^{(i) } \ominus \mathcal{B}(0,\epsilon)$.
\end{lemma}

\begin{proof} First, the abstraction function $\gamma_{\eta}$ satisfies $\gamma_{\eta}(\Theta) \subset (\Theta \oplus \mathcal{B}(0,\eta/2)) \quad  \forall \Theta \subset \mathbb{R}^d$
from which we obtain
    \begin{equation*}
        \gamma_{\eta^{(i)}} \left ( \Theta_{\text{safe}}^{(i)} \ominus \mathcal{B}(0,\delta) \right ) \subset \left ( \Theta_{\text{safe}}^{(i)} \ominus \mathcal{B}(0,\delta) \right ) \oplus \mathcal{B} \left ( 0, \frac{\eta^{(i)}}{2} \right ).
    \end{equation*}
    Also, we have
    \begin{equation*}
        \begin{aligned}
            & \left ( \Theta_{\text{safe}}^{(i)} \ominus \mathcal{B} (0,\delta) \right ) \oplus \mathcal{B} \left ( 0, \frac{\eta^{(i)}}{2} \right ) \oplus \mathcal{B} (0,\epsilon)  \\
            = & \left ( \Theta_{\text{safe}}^{(i)} \ominus \mathcal{B}(0,\delta) \right ) \oplus \mathcal{B} \left ( 0, \epsilon + \frac{\eta^{(i)}}{2} \right ) \\
            \subset & \left ( \Theta_{\text{safe}}^{(i)} \ominus \mathcal{B}(0,\delta) \right ) \oplus \mathcal{B} (0,\delta) = \Theta_{\text{safe}}^{(i)},
        \end{aligned}
    \end{equation*}
    where the first equation uses the definition of the Minkowski sum, and the second relationship between the sets comes from $\delta > \epsilon + \eta^{(i)} / 2$. Since $\Theta_{\text{safe}}^{(i)} \ominus \mathcal{B} (0,\epsilon) $ is the largest set in $X$ that satisfies $X \oplus \mathcal{B} (0, \epsilon) = \Theta_{\text{safe}}^{ (i)}$, the following holds,
    \begin{equation*}
        \begin{aligned}
            \gamma_{\eta^{(i)}} ( \Theta_{\text{safe}}^{(i)} \ominus \mathcal{B} (0,\delta))\subset & \left ( \Theta_{\text{safe}}^{(i)} \ominus \mathcal{B}(0,\delta) \right ) \oplus \mathcal{B} \left ( 0, \frac{\eta^{(i)}}{2} \right ) \\
            \subset & \Theta_{\text{safe}}^{(i)} \ominus \mathcal{B} (0,\epsilon),
        \end{aligned}
    \end{equation*}
    and the statement has been proven.
\end{proof}

\begin{proof} 
     [Theorem~\ref{thm:relmaxinv1}] Let $\delta$ be larger than $\epsilon + \eta^{(i)} / 2$. From the statement of the Theorem, the following holds
    \begin{align}
        & \xi_{j}^{(i)} (t) \xrightarrow[(i) , \eta^{(i)} ]{\mu_j^{(i)} (t)}  \xi_j^{(i)} (t+1) & \forall t  \in \mathbb{N}_0 \nonumber \\
        & \xi_j^{(i)} (t) \in \gamma_{\eta^{(i)}}(\Theta_{\text{safe}}^{(i) } \ominus \mathcal{B}(0,\delta)) & \forall t \in \mathbb{N}_0 \label{th1:eq1}.
    \end{align}
        
     Assume that $\theta_j^{(i)} (0)$ satisfies $\| \theta_j^{(i)} (0) - \xi_j^{(i)} (0) \| \leq \epsilon$, and let $\theta_{j}^{(i)} (t)$ be the state trajectory of $S_j^{(i)}$ starting from $\theta_j^{(i)} (0)$ under input trajectory $\mu_j^{(i)} (t)$. By the assumption that $S^{(i)}$ and $S_{\eta}^{(i)}$ are $\epsilon$-approximately bisimilar, $\| \theta_j^{(i)} (t) - \xi_j^{(i)} (t) \|$ is always smaller than or equal to $\epsilon$. Moreover, from Lemma~\ref{lem:invlemma1} and \eqref{th1:eq1}, we have $\xi_j^{(i)} (t) \in \Theta_{\text{safe}}^{(i)} \ominus \mathcal{B}(0,\epsilon)$. Therefore, $\theta_j^{(i)}$ belongs to  $( \Theta_{\text{safe}}^{(i)} \ominus \mathcal{B}(0,\epsilon)) \oplus \mathcal{B}(0,\epsilon)  = \Theta_{\text{safe}}^{(i)}$ for any $t$ in $\mathbb{N}_0$, which proves the theorem.
\end{proof}

\subsection*{Proofs for Section~\ref{sec:invsetmethod} :}

\begin{proof}[Theorem \ref{theo:Xcyc_inv}]
    We first show that $\bm{X}_{\text{inv}}$ is a subset of $\bm{X}_{\text{safe}}$. Let $\bm{x} := (x^{(1) \top}, \ldots, x^{(g)\top})^{\top}$ be an element of $\bm{X}_{\text{inv}}$. By the definition of $\bm{X}_{\text{inv}}$, there exists a vector $(\beta_1^{(1) \top},\ldots, \beta_{n^{(g)}}^{(g) \top})^{\top} \in \Omega$ such that $x^{(i)} = \sum_{j=1}^{n^{(i)}} \Phi_j^{(i)} ( \beta_j^{(i)} )$ for all $i$ in $[g]$. Then, for any $k \in [K^{(i)}] \setminus \mathcal{I}_{\text{safe}}^{(i)}$, we can show 
    \begin{equation} \label{eq:thm2eq1}
        x_{m,\tau,k}^{(i)} = \sum_{j=1}^{n^{(i)}} \sum_{l: \tilde{\nu}_{j,l}^{(i)} = \nu_{m,\tau,k}^{(i)}} [ \beta_j^{(i)} ]_l = 0,
    \end{equation}
    since all the nodes $\tilde{\nu}_{j,l}^{(i)}$ belong to $V_{\text{safe}}^{(i)}$. 
    
    Also, the following equation holds,
    \begin{equation*}
        \begin{aligned}
            \sum_{\tau=0}^{\overline{\tau}_{i,m}} \sum_{k=1}^{K^{(i)}} x_{m,\tau,k}^{(i)} = &  \sum_{\tau=0}^{\overline{\tau}_{i,m}} \sum_{k=1}^{K^{(i)}}  \sum_{j=1}^{n^{(i)}} \sum_{l: \tilde{\nu}_{j,l}^{(i)} = \nu_{m,\tau,k}^{(i)}} [ \beta_j^{(i)} ]_l  \\
            = & \sum_{j=1}^{n^{(i)}} \left ( \sum_{\tau=0}^{\overline{\tau}_{i,m}} \sum_{k=1}^{K^{(i)}} \sum_{l: \tilde{\nu}_{j,l}^{(i)} = \nu_{m,\tau,k}^{(i)}} [ \beta_j^{(i)} ]_l  \right ) \\
            = &  \sum_{j=1}^{n^{(i)}} \sum_{l : \tilde{\mu}_{j,l}^{(i)} = m } [ \beta_j^{(i)} ]_l = \sum_{j=1}^{n^{(i)}} H_{m,0} (\beta_j^{(i)}).
        \end{aligned}
    \end{equation*}
    Therefore, we obtain the following
    \begin{equation} \label{eq:thm2eq2}
        \sum_{i=1}^{g} p_{m}^{(i)} \sum_{\tau=0}^{\overline{\tau}_{i,m}} \sum_{k=1}^{K^{(i)}} x_{m,\tau,k}^{(i)} = \sum_{i=1}^{g} p_{m}^{(i)} \sum_{j=1}^{n^{(i)}} H_{m,0} (\beta_j^{(i)}).
    \end{equation}
    From \eqref{eq:cycassaggconst} and  \eqref{eq:thm2eq2}, the following holds
    \begin{equation} \label{eq:thm2eq3}
        \underline{P}_m \leq \sum_{i=1}^{g} p_{m}^{(i)} \sum_{\tau=0}^{\overline{\tau}_{i,m}} \sum_{k=1}^{K^{(i)}} x_{m,\tau,k}^{(i)} \leq \overline{P}_m \quad \forall m \in [M].
    \end{equation}
    By \eqref{eq:thm2eq1} and \eqref{eq:thm2eq3}, $(x^{(1) \top},\ldots,x^{(g) \top})^\top$ belongs to $\bm{X}_{\text{safe}}$.
    
    Next, we show the recurrence property of $\bm{X}_{\text{inv}}$. Let $\tilde{\bm{u}}$ be the input corresponding to 1-step of circular shift from $(\beta_1^{(1) \top},\ldots,\beta_{n^{(g)}}^{(g) \top})^\top$, and $\tilde{\bm{x}} := (\tilde{x}^{(1) \top},\ldots,\tilde{x}^{(g) \top})^\top$ be the corresponding next state from $\bm{x}$ under $\tilde{\bm{u}}$ (i.e., $\tilde{\bm{x}} = A \bm{x} + B \tilde{ \bm{u}}$). Then,  $\tilde{x}^{(i)}$ is equal to $\sum_{j=1}^{ n^{(i)} } \Phi_j^{(i)}  ( \Psi_{l_j^{(i)}} \beta_j^{(i)})$ for any $i$ in $[g]$. To show that $\tilde{\bm{x}}$ is an element of $\bm{X}_{\text{inv}}$, we show that $( (\Psi_{l_1^{(1)}} \beta_1^{(1)})^\top,\ldots, (\Psi_{l_{n^{(g)}}^{(g)}} \beta_{ n^{(g)} }^{(g)})^\top )^\top$ belongs to $\Omega$.
    
    First, from $\bm{1}^{\top}  \Psi_{l_j^{(i)}} = \bm{1}^{\top}$, the following is easily shown
    \begin{equation} \label{eq:thm2eq4}
        \sum_{j=1}^{n^{(i)}} \bm{1}^{\top}  \Psi_{l_j^{(i)}} \beta_j^{(i)}  = \sum_{j=1}^{n^{(i)}} \bm{1}^{\top} \beta_j^{(i)} = N^{(i)}. 
    \end{equation}
    Also, using $ H_{m,q'} ( \Psi_{l_j^{(i)}} \beta_j^{(i)} ) =  H_{ m, \tilde{q} } ( \beta_j^{(i)} ) $ where $ \tilde{q} $ is $ q'+1 \; (\text{mod} \; l_j^{(i)})$, we can show the following
\begin{equation*}
    \min_{q' \in [ l_j^{(i)} ] } H_{m,q'} \left ( \Psi_{l_j^{(i)}} \beta_j^{(i)} \right ) = \min_{q' \in [ l_j^{(i)} ] } H_{m,q'} \left ( \beta_j^{(i)} \right ).
\end{equation*}
Hence, from \eqref{eq:safecycasscond1}, the following holds for any $m \in [M]$
\begin{equation} \label{eq:thm2eq5}
      \sum_{i=1}^{g} p_{m}^{(i)} \sum_{j=1}^{ n^{(i)}}  \min_{q' \in [l_j^{(i)}]} H_{m,q'} \left ( \Psi_{l_j^{(i)}} \beta_j^{(i)} \right ) \geq \underline{P}_m.
\end{equation} 
Similarly, the following holds for any $m \in [M]$
\begin{equation} \label{eq:thm2eq6}
      \sum_{i=1}^{g} p_{m}^{(i)} \sum_{j=1}^{ n^{(i)}}  \max_{q' \in [l_j^{(i)}]} H_{m,q'} \left ( \Psi_{l_j^{(i)}} \beta_j^{(i)} \right ) \leq \overline{P}_m .
\end{equation}
 By \eqref{eq:thm2eq4}, \eqref{eq:thm2eq5}, \eqref{eq:thm2eq6}, 
  $( (\Psi_{l_1^{(1)}} \beta_1^{(1)})^\top,\ldots, (\Psi_{l_{n^{(g)}}^{(g)}} \beta_{ n^{(g)} }^{(g)})^\top )^\top$ belongs to $\Omega$. Therefore, $\tilde{\bm{x}}$ belongs to $\bm{X}_{\text{inv}}$ which shows the recurrence property of $\bm{X}_{\text{inv}}$.
\end{proof}

\vskip 0pt plus -1fil






\end{document}